\documentclass[prb,aps,showpacs,superscriptaddress,twocolumn]{revtex4}
\usepackage{amsmath}
\usepackage{amssymb,color}
\usepackage {graphicx}

\newcommand{\beq}{\begin{equation}}
\newcommand{\eeq}{\end{equation}}
\newcommand{\bea}{\begin{eqnarray}}
\newcommand{\eea}{\end{eqnarray}}

\newcommand{\BFCA}{Ba(Fe$_{1-x}$Co$_x$)$_2$As$_2$}

\newcommand{\KFA}{KFe$_2$As$_2$~}
\newcommand{\BFAP}{BaFe$_2$(As$_{1-x}$P$_x$)$_2$}

\newcommand{\BKFA}{Ba$_{1-x}$K$_x$(FeAs)$_2$}

\begin{document}

\title{On the gap symmetry in KFe$_2$As$_2$ and $\cos 4 \theta$ gap component in LiFeAs }
\author{S.~Maiti}
\affiliation{Department of Physics, University of Wisconsin,
Madison, Wisconsin 53706, USA}
\author{M.M.~Korshunov}
\affiliation{L.V. Kirensky Institute of Physics, Siberian Branch
of Russian Academy of Sciences, 660036 Krasnoyarsk, Russia}
\affiliation{Siberian Federal University, Svobodny Prospect
79,660041 Krasnoyarsk, Russia.}
\author{A.V.~Chubukov}
\affiliation{Department of Physics, University of Wisconsin,
Madison, Wisconsin 53706, USA}

\date{\today}


\begin{abstract}
We revisit the issue of the gap symmetry in \KFA, which  is an
Fe-pnictide superconductor with only hole pockets. Previous
theoretical studies mostly argued for a $d-$wave gap in \KFA since
transport and thermodynamic measurements point to the presence of
the gap nodes. However, a $d-$wave gap is inconsistent with recent
laser-based angle-resolved photoemission measurements. We propose
the scenario for a nodal $s-$wave superconductivity induced by a
non-magnetic intra-band and inter-band interactions between
fermions near hole pockets. The superconducting gap changes sign
between the hole pockets and has $\cos {4\theta}$ angular
dependence and accidental nodes on one or several hole pockets. We
argue that strong angle dependence is the consequence of
near-degeneracy between inter-pocket and intra-pocket interaction
on the hole pockets.  We also analyze $\cos {4\theta}$ angular
dependence of the gap in other Fe-pnictides and compare theoretical results
with the photoemission experiments of LiFeAs.
\end{abstract}

\maketitle

\section{Introduction}

One of the most interesting features in the rapidly growing family
of the Fe-based superconductors(FeSCs) is the manifestation of
different gap structures in the superconducting (SC) state which
may potentially indicate different gap symmetries in different
materials~\cite{review}. An important step towards identifying the
gap symmetry is to establish whether or not the system shows nodal
behavior and whether or not the nodes are symmetry-related. In the
weakly and moderately doped FeSCs, which have electron-like and
hole-like pockets, there is a strong experimental evidence
~\cite{review} and general agreement amongst theoretical
approaches~\cite{fRG_wdFeSC_1,fRG_wdFeSC_2,RPA_wdFeSC_1,RPA_wdFeSC_2,aRG_wdFeSC}
that the gap symmetry is of $s^{\pm}$ type with opposite sign of
the gap on the electron and hole pockets. Angle-Resolved
Photo-Emission Spectroscopy (ARPES) on optimally doped
\BKFA~\cite{ARPES_BKFA_1, ARPES_BKFA_2}, \BFCA~\cite{ARPES_BFCA}
and undoped LiFeAs\cite{ARPES_LiFeAs_1,ARPES_LiFeAs_2} has
identified nodeless gaps on the hole pockets ruling out non-s-wave
gap symmetry, up to certain exceptions\cite{comm}. Thermodynamic
measurements on these materials~
\cite{Transport_BKFA,Transport_BFCA,Transport_LiFeAs} show
nodeless behavior consistent with $s-$wave gap symmetry. In some
other compounds (such as \BFAP) electronic transport and
thermodynamic measurements\cite{Transport_BFAP} reveal the
presence of nodes and at the same time ARPES\cite{ARPES_BFAP}
clearly rules out nodes on hole pockets. This is still consistent
with $s^{\pm}$ symmetry provided that the gap nodes are accidental
and reside on the electron pockets. The reasoning behind the
existence of no-nodal and nodal $s^{\pm}$ gap structures is
specific to the Fermi surface (FS) geometry  in FeSCs and is
related to the interplay between inter-pocket and intra-pocket
electron(e)-hole(h) couplings ($u_{eh}$ and $u_{ee}$, $u_{hh}$,
respectively). When the interactions are even slightly
anisotropic, a larger inter-pocket coupling
($u_{eh}^2>u_{ee}u_{hh}$)  leads to a nodeless structure while a
smaller inter-pocket coupling ($u_{eh}^2<u_{ee}u_{hh}$) leads to
accidental nodes on electron pockets\cite{aRG_wdFeSC,LAHA_long}.

This reasoning, however, does not apply to FeSCs with only one
kind of carriers, either holes or electrons. In this article we
focus on systems with only hole pockets.  The primary example here
is \KFA which is at the end point of the family of hole-doped
\BKFA. Synchrotron-based ARPES measurements on this
material~\cite{yoshida,ding,evtush} show two distinct hole pockets
centered at $\Gamma$ point (${\bf k} = (0,0)$) and small hole
blades near $(\pi,\pi)$ point, but no electron pockets. Recent
laser-based ARPES measurement~\cite{shin} resolved three hole
pockets near $\Gamma$, of which, the inner and the middle ones,
two have orbital content $d_{xz}/d_{yz}$ and the third, outer
pocket is $d_{xy}$. The electronic structure with 3 hole pockets
at $\Gamma$ and hole blades at the corners of the Brillouin zone
(BZ) is consistent with DFT band structure calculations for this
material~\cite{band_theory} and with the trends observed in ARPES
studies of \BKFA with increasing $x$ (Ref. \onlinecite{evtush2}).

Penetration depth and transport measurements of \KFA
\cite{Transport_KFA} show linear in $T$  behavior, similar to that
in \BFAP. There are two existing theoretical scenarios for this
behavior. One is that the gap has $d_{x^2-y^2}$ symmetry, induced
primarily by $2k_F$ intra-pocket interaction within the largest
hole pocket. In this case, the gap has nodes along $k_x =k_y$
which naturally explains the observed linear in $T$ behavior.
Another scenario is that the magnetically enhanced electron-hole
interaction still gives rise to $s^{\pm}$
superconductivity~\cite{kuroki}, despite the electron states being
gapped. Within this scenario, the observed linear in $T$
dependencies are either due to the smallness of the gap on one of
hole pockets or, potentially, due to the presence of horizontal
line nodes at some $k_z$ on the other hole pocket. Functional
renormalization group (fRG) study\cite{fRG_KFA} found $d-$wave as
a clear winner, while FS-restricted RPA-type spin-fluctuation (SF)
analysis~\cite{kuroki} and the analytical study in which the
interactions are approximated by their lowest angular harmonics
(LAH)~\cite{LAHA_long} have found that $s-$wave and $d-$wave
pairing amplitudes have near-equal strength, primarily because
$2k_F$ for the largest hole FS is not far from the distance
between hole and the would-be electron pockets.

Because $d-$wave gap naturally explains  the linear in $T$
behavior it had been generally considered as the most plausible
gap structure. This further  raised speculations about potential
time-reversal symmetry breaking $s+id$ pairing state in
Ba$_{1-x}$K$_x$Fe$_2$As$_2$  at $x \leq 1$ \cite{s+id} However,
the neutron scattering analysis of the vortex lattice in
\KFA~\cite{vortex} and, particularly, the laser-based ARPES
measurements of the gaps along the three hole FSs at $\Gamma$
point~\cite{shin} cast strong doubts that the gap is $d-$wave. The
ARPES data show that:  (i)  the measured gap is definitely the
smallest at the outer pocket, while all theoretical calculations
point out~\cite{kuroki,fRG_KFA,LAHA_long} that it should be the
largest, if the gap is $d-$wave; (ii) the gap along the smallest
(inner) hole pocket has angular dependence but no nodes, and (iii)
the gap along the middle hole pocket has accidental nodes. All
three results are inconsistent with the $d-$wave gap.  The data
are partly consistent with the theoretical prediction for $s-$wave
gap~\cite{kuroki,LAHA_long} in that it is the smallest on the
outer pocket.  However, the substantial angular variation of the
gaps and the nodal behavior of the gap on the middle hole FSs is
inconsistent with the angle-independent $s-$wave gap obtained in
the LAH approximation~\cite{LAHA_long} and also inconsistent with
the SF analysis according to which $s-$wave gap has no nodes
except at a particular $k_z$ where the gap on the middle hole
pocket just vanishes. Experimentally, laser-based ARPES  measures
the gap averaged over some range of $k_z$~\cite{shin}, which
should wipe out such horizontal nodes.

In this article we argue that the superconducting state with an
$s-$wave gap symmetry {\it and} nodes appears quite naturally when
only hole pockets are present if the interaction between fermions
is the largest at a small momentum transfer. This is the case when
spin and charge fluctuations are not strong, and the effective
pairing interaction is well approximated by the first-order term,
which is the combination of Hubbard and Hund intra- and
inter-orbital interactions, dressed up by `coherence factors'
which hybridize electron orbitals and produce bands and
consequently hole pockets. We consider only direct $Fe-Fe$
interaction and neglect the induced interaction via a pnictide. In
this situation, to treat interactions adequately one should move
to the unfolded BZ, in which the hole blades are at $(\pi,0)$ and
symmetry-related points, and the  outer $d_{xy}$ hole pocket is at
$(\pi,\pi)$.

When the interaction in the unfolded zone is peaked at a small
momentum transfer,  the $(\pi,\pi)$ hole pocket and $(\pi,0)$ hole
blades are not overly relevant, and the physics is determined by
intra-pocket and inter-pocket interactions for the two hole
pockets centered at $\Gamma$ point. For strictly angle-independent
interactions, superconductivity is only possible when the
inter-pocket coupling  exceeds the intra-pocket one. Then the
system develops an $s^{\pm}$ gap which changes sign between the
two hole pockets.  We show, however, that, once we allow the
interaction to have some angular dependence, superconductivity
develops even when intra-pocket coupling is larger. Moreover, when
intra-pocket and inter-pocket interactions are of near-equal
strength, which is the case for \KFA because both hole FSs are
small and centered at the same point, the angular component of
intra-pocket interaction is enhanced in a resonance-type fashion,
and the SC gap acquires strong variations along the hole FSs and
accidental nodes, even if this angular component is small. Such
resonant enhancement of the angular-dependent component of the gap
has been previously found~\cite{CVV} for the case when both hole
and electron pockets are present and the dominant inter-pocket
interaction is between hole and electron pockets.  Here we apply
the same reasoning to the case when inter-pocket interaction is
between the two hole pockets at the $\Gamma$ point.

For completeness, we
 also consider in some detail $\cos 4 \theta$ gap components on hole pockets
 in other FeSCs, which have both hole and electron FSs, and discuss the interplay
  between $\cos 4 \theta$ and $\cos 2 \theta$ gap variations along electron pockets.
 In this context we compare theoretical results with recent
  ARPES  measurements  of gap variations in LiFeAs along both hole and electron FSs~\cite{sergei,ding_LiFeAs}.
  We argue, in particular,  that the data unambiguously show the presence  of
  $\cos {4\theta}$ {\it and}  $\cos {2\theta}$  oscillations on electron FSs.
 We argue that the sign of $\cos 2 \theta$ term is consistent with the theoretical prediction for the case
  when the pairing interaction is predominantly between electron and hole pockets.

\section{The approach}

We consider intra-pocket and inter-pocket pairing interactions
between low-energy hole-like fermions which we label as $h_i$
where $i$ specifies the pocket. The physics we consider is not
overly sensitive to $k_z$ variations of the electronic dispersion,
and we restrict ourselves to a pure 2D model. The interactions
$U_{h_i,h_j} ({\bf k}_{F_i}, {\bf k}_{F_j})$, generally depend on
the angles along hole FSs.  By general reasons \cite{LAHA_long}
the angle dependencies of the
 interactions $U_{h_i,h_j} ({\bf k}_{F_i}, {\bf
k}_{F_j})$ can be expanded in powers of $\cos{4n \theta_i}$,
$\cos{4n \theta_j}$, where $\theta_{i,j}$ are the angles along the
corresponding FS~\cite{aRG_wdFeSC,LAHA_long}. The first term in
the series is a constant $U_{h_i,h_j}$ for $s-$wave pairing and
$\cos{2 \theta_i} \cos{2\theta_j}$ for $d-$wave pairing. The
$d-$wave pairing has been analyzed in detail in Ref.
\onlinecite{LAHA_long}, here we consider $s-$wave pairing.

In Ref. \onlinecite{LAHA_long} only the constant $s-$wave terms
had been kept. Accordingly, $s-$wave  gaps $\Delta_{i}$ were
angle-independent.  Here we go a step further and include into
consideration the sub-leading $\cos 4 \theta_{i,j}$ and, later,
also $\cos 8 \theta_{i,j}$ terms in $U_{h_i,h_j} ({\bf k}_{F_i},
{\bf k}_{F_j})$. This gives rise to $\cos 4 \theta_i$ and $\cos 8
\theta_i$ angle dependent terms in the gaps $\Delta_i$. We do not
consider higher-order harmonics as measured gaps are well fitted
by the $\cos 4 \theta_i$ and $\cos 8 \theta_i$ forms~\cite{shin}.

The magnitude of angle-dependent terms in the interactions
$U_{h_i,h_j} ({\bf k}_{F_i}, {\bf k}_{F_j})$ depends on the
structure of the hopping integrals in the orbital
basis~\cite{graser} and can be material-dependent. Still, ARPES
data on  weakly/moderately doped \BKFA  show almost no angle
variation on the hole gaps which most likely indicates that the
angle-dependent terms in $U_{h_i,h_j} ({\bf k}_{F_i}, {\bf
k}_{F_j})$ in K-doped 122 systems are weak~\cite{ARPES_BFAP}. Naively,
one could then expect little angle dependence of $s-$wave gaps in
\KFA as well. We show, however, that this is not the case if  the
pairing interaction in \KFA involves the two hole pockets centered
at $\Gamma$. Then the angular dependent $\cos 4\theta$ term in the
gap is enhanced in a resonance-like fashion  and may give rise to
accidental nodes.

That an $s-$wave gap  on the hole pocket can have nodes due to
strong $4\theta$ variations that has been found
before\cite{graser2} in the study of a 3D band structure of \BKFA
where electron pockets where still present and the pairing was
driven by large momentum scattering enhanced by spin fluctuations.
In that work, the nodes were only present for a particular $k_z$.
In our case, the gap has vertical line nodes, present at any
$k_z$.

To make the presentation more transparent and physically
insightful, we first consider analytically in the next section the
case of just two hole pockets $h_1$ and $h_2$ at $(0,0)$,  include
the $\cos 4 \theta$ dependence of the intra-pocket $h_1-h_2$
interaction and show how accidental nodes appear  on either both
or one of these pockets already for weakly angle-dependent
interaction. We then consider in Sec. \ref{sec:full} the full
model with 3 hole pockets, use as inputs the interactions
$U_{h_i,h_j} ({\bf k}_{F_i}, {\bf k}_{F_j})$ obtained from the
underlying 5-orbital model, fit all interactions by the first
three angular harmonics (a constant, $\cos 4 \theta_i$, and $\cos
8 \theta_i$ terms) and solve $9\times 9$ matrix gap equation. We
show that intra-pocket and inter-pocket interactions involving
$h_1$ and $h_2$ pockets are almost identical, and the angular
dependent components of the interactions are rather weak.
Nevertheless, the solution of $9\times 9$ matrix gap equation
shows that the $s-$wave gap has accidental nodes on at least one
hole pocket. This analysis also confirms that the gap at
$(\pi,\pi)$ pocket is rather small, i.e., the pairing is
predominantly determined by the interactions between fermions near
the $\Gamma$ point.

In Sec.\ref{sec:sign} we consider relative phases of $\cos 4
\theta$ terms on different hole FSs and argue that the sign of the
$\cos 4 \theta$ component of the gap on a given FS is at least
partly determined by its shape. In this section we also discuss
the situation in another Fe-pnictide, LiFeAs,  in which $\cos 4
\theta$ variations have been observed on both hole and electron
FSs~\cite{sergei,ding_LiFeAs,rost}. We argue that the data also
show the presence of $\cos 2 \theta$ variations along electron
FSs, predicted by the theory.

\section{The model with two hole pockets at $(0,0)$}
\label{sec:equiv}

Like we said above, we keep $\cos 4 \theta$  angular dependence in
the inter-pocket interaction  $U_{h_1 h_2}$ and approximate  the
intra-pocket $U_{h_1 h_1}$ and $U_{h_2 h_2}$ by constants, i.e.,
set \bea\label{interactions_toy}
U_{h_1h_1}(k,p)&=&U_{11};~~U_{h_2h_2}(k,p)=U_{22}\nonumber\\
U_{h_1h_2}(k,p)&=&U_{12} \left (1 +\alpha_{12} \cos
4\theta_k+\alpha_{21} \cos 4\theta_p \right) \eea where $\theta$
are measured from the $x$ axis.

The linearized BCS gap equation that one needs to solve is
\bea\label{gap_eq_toy}
 \Delta_{h_1}(p)&=&-L N_{F_1} \int_0^{2\pi}\frac{d
\theta_k}{2\pi}
U_{h_1h_1}(p,k)\Delta_{h_1}(k)-\nonumber\\
&&L N_{F_2} \int_0^{2\pi}\frac{d
\theta_k}{2\pi} U_{h_1h_2}(p,k)\Delta_{h_2}(k)\nonumber\\
\Delta_{h_2}(p)&=&-L N_{F_1} \int_0^{2\pi}\frac{d \theta_k}{2\pi}
U_{h_2h_1}(p,k)\Delta_{h_1}(k)-\nonumber\\
&&L N_{F_2} \int_0^{2\pi}\frac{d \theta_k}{2\pi}
U_{h_{2}h_2}(p,k)\Delta_{h_2}(k)\nonumber\\
\eea
where $L=log\frac{\Lambda}{T_c}$ and $N_{F_{1,2}}$ are the
densities of states. For  the interaction of Eq.
(\ref{interactions_toy}), the gap structure is
\bea\label{gaps_toy} \Delta_{h_1}(p)&=&\Delta_1\left( 1+ r_1 \cos
4\theta_p\right)\nonumber\\
\Delta_{h_2}(p)&=&\Delta_2\left( 1+ r_2 \cos 4\theta_p\right) \eea

\subsection{Equivalent hole pockets}

We first consider the case when the densities of states on the two
hole pockets are the same, $N_{F_1}=N_{F_2}\equiv N_F$, and then
extend the consideration to $N_{F_1} \neq N_{F_2}$. For the first
case, it is natural to take $U_{11} = U_{22} = U$ and $\alpha_{12}
= \alpha_{21} = \alpha$, and introduce dimensionless couplings
$u_{12} = N_F U_{12}$ and $u = N_F U$. For positive $u_{ab}$,
which we consider, the solution of the gap equation is
$\Delta_{h_1}=-\Delta_{h_2}$ with $r_1=r_2=u_{12}\alpha L$ and it
develops at

\beq\label{L}
L=\frac{-(u-u_{12})+\sqrt{(u-{u_{12}})^2+2(u_{12}\alpha)^2}}{(u_{12}\alpha)^2}
\eeq

Observe that, for any $\alpha \neq 0$, $L >0$  (i.e., $T_c >0$) no
matter what the signs of $u_{12} -u$ and $\alpha$ are (c.f.
Ref.\onlinecite{CVV}). Observe also that  this $s^{\pm}$ gap
structure with opposite signs of the gaps on different hole
pockets  is different form that proposed in
Ref.\onlinecite{kuroki} where the gaps on hole pockets have the
same sign and that on the would-be electron pockets have opposite
sign.

\begin{figure}[t]
$\begin{array}{cc}
\includegraphics[width=1.8in]{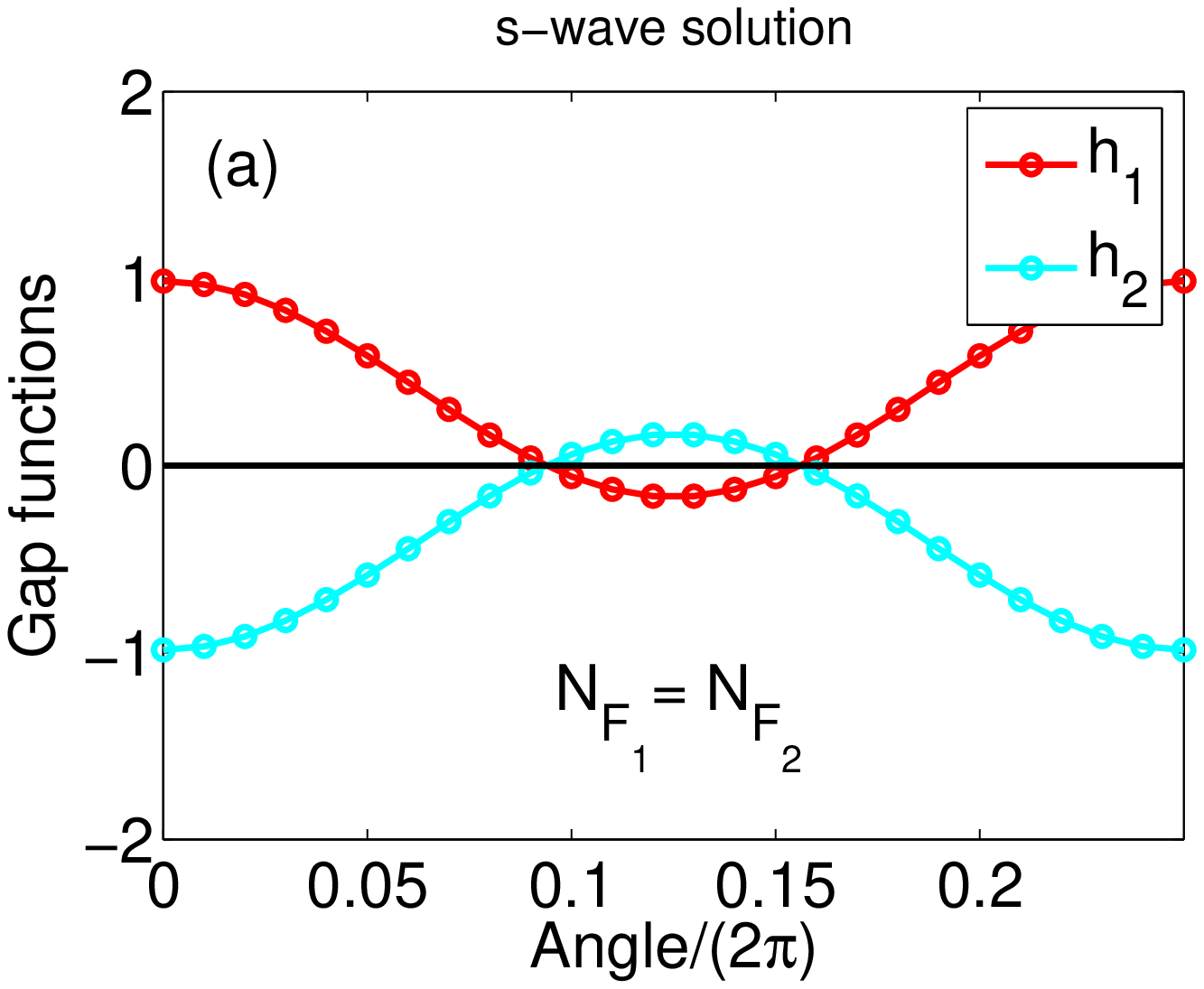}&
\includegraphics[width=1.8in]{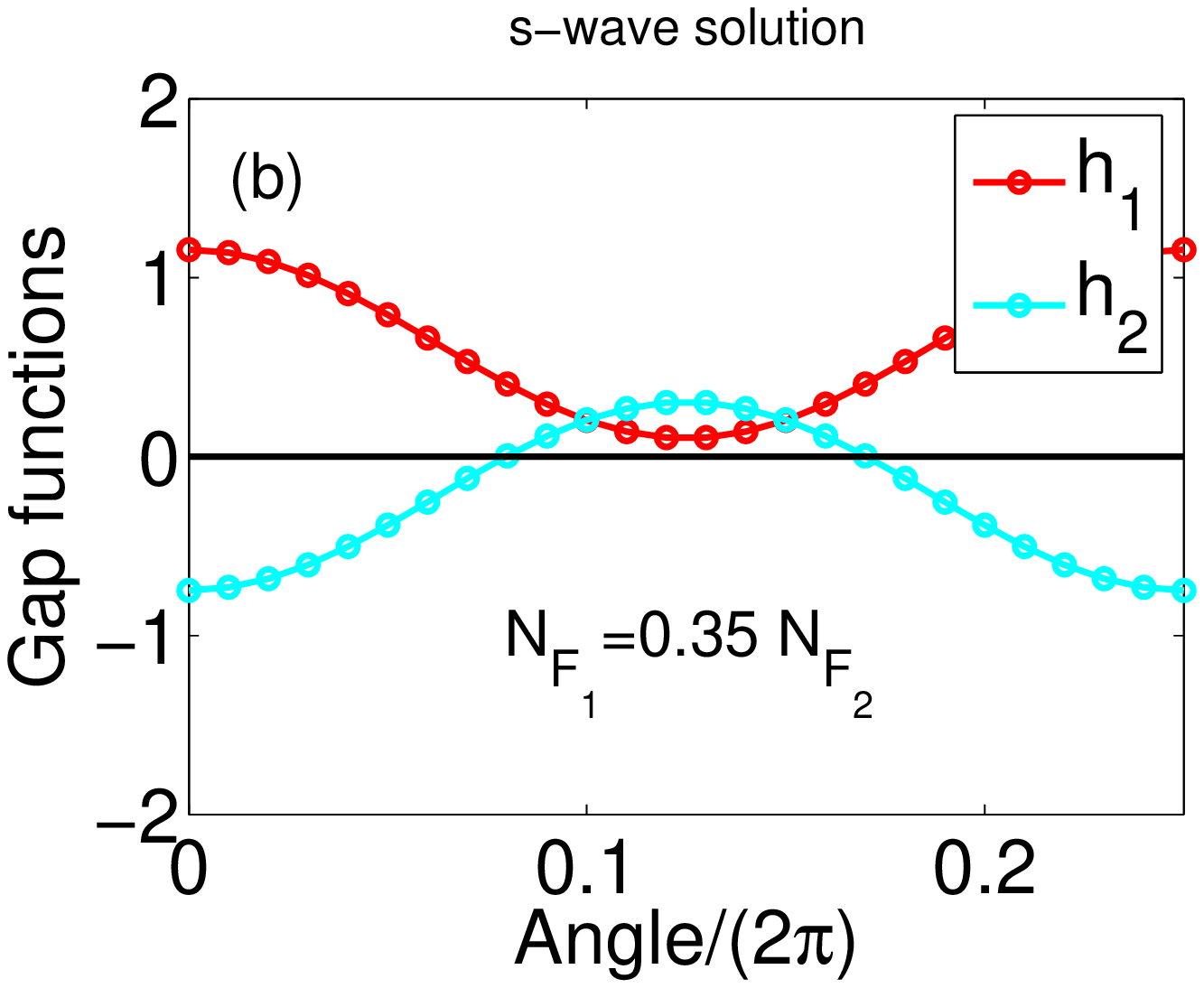}\\
\includegraphics[width=1.8in]{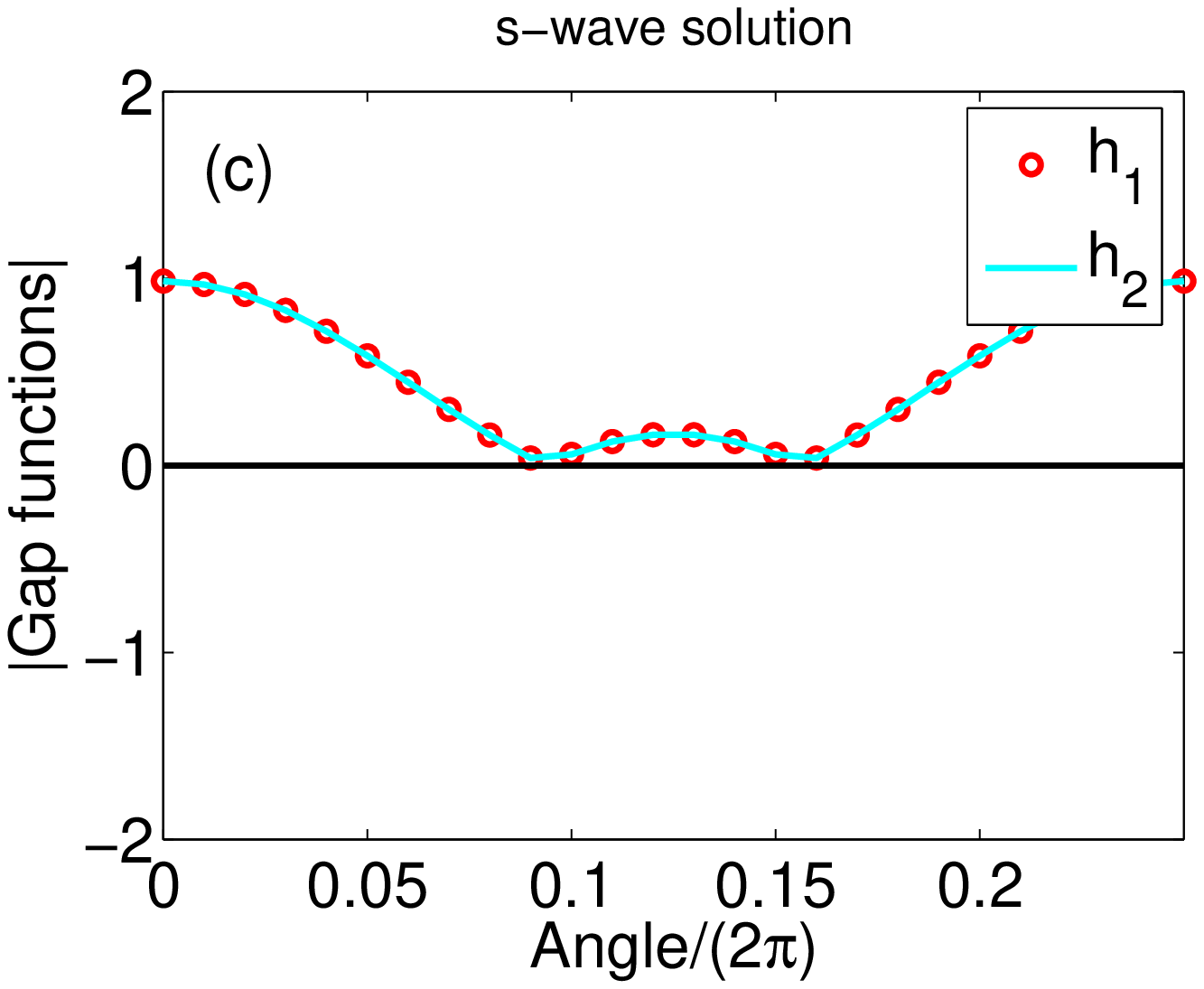}&
\includegraphics[width=1.8in]{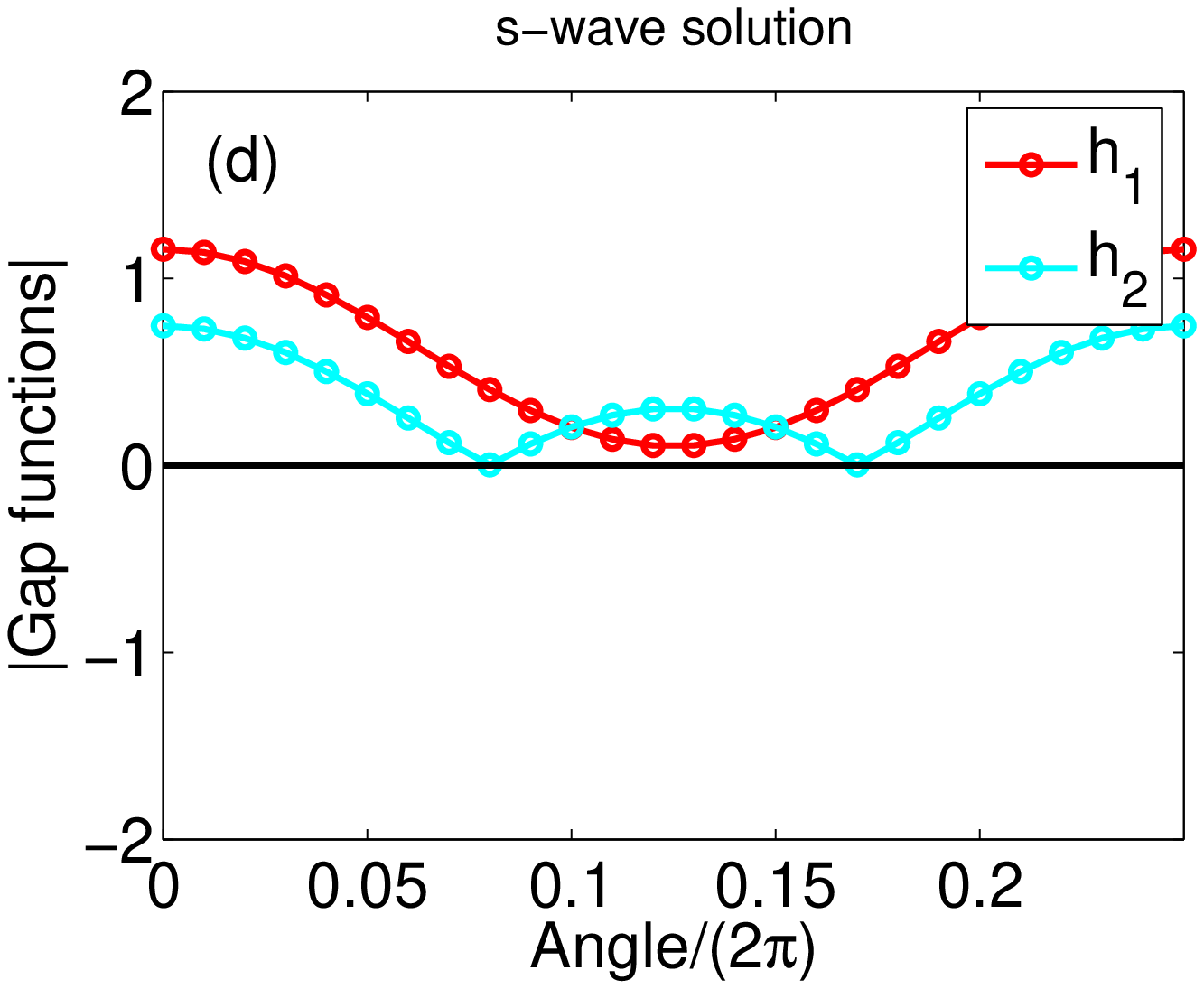}
\end{array}$
\caption{\label{fig:DOS} (a) The gap structure obtained for equal
density of states ($N_{F_1}=N_{F_2}$) on the two hole pockets at
$\Gamma$. We used $U_{11}=U_{22}=U_{12}$ and $\alpha=0.1$. The gap
satisfies $\Delta_{h_1}=-\Delta_{h_2} = \Delta(1+ \sqrt{2} \cos 4
\theta)$ and has nodes on both FSs. (b)The same as (a), but with
$N_{F_1}= 0.35 N_{F_2}$. Now the two gaps are non-equivalent and
the gap on the inner hole FS has angle variation but no nodes. For
comparison with ARPES, in panels  (c) and (d) we plot the absolute
values of the gaps from panels (a) and (b), respectively.}
\end{figure}

One can further discuss the characteristics of the solution  by
looking into different parameter regimes. For small $\alpha$ and
$u>u_{12}$, we have $r_1=\frac{2(u-u_{12})}{|u_{12}\alpha |}$ and
$L=\frac{2(u-u_{12})}{(u_{12}\alpha)^2}$. For small $\alpha$ this
yields the solution with nodes ($r_1>1$), although $T_c$ is quite
small. For $u<u_{12}$, we have $r_1=\frac{|u_{12}\alpha
|}{|u-u_{12}|}<1$ and $L=\frac{1}{|u-u_{12}|}$. For small
$\alpha$, the gap is nodeless, and $T_c$ is larger than in the
previous case. The most interesting case, relevant to \KFA, is the
one with $u=u_{12}$ (because the two FSs are centered at the same
$\Gamma$ point, intra-pocket and inter-pocket interactions are
undistinguishable). In this case we get $r_1=\sqrt{2}$ independent
of $\alpha$, indicating that nodes will arise even for a very weak
angular dependence of the interaction~\cite{CVV}. This is what we
mean by a resonant enhancement of the angular dependence of the
interactions. Fig. \ref{fig:DOS}(a) shows the gap structure
obtained for this case.

\subsection{Non-equivalent hole pockets}

We next consider the case of non-equivalent hole pockets, when
$N_{F_1} \neq N_{F_2}$ and, generally, $U_{11} \neq U_{22}$. Both
lead  to $u_{11} = N_{F_1} U_{11}$ being different from $u_{22} =
N_{F_2} U_{22}$. In principle, $\alpha_{12}$ is also different
from $\alpha_{21}$, but since we keep $\alpha$ small, a inequality
of $\alpha_{12,21}$ is not relevant and we continue to treat them
equal ($ \alpha_{12} = \alpha_{21} =\alpha$).

There are two issues that one need to consider for non-equivalent
pockets -- whether the solution exists for all parameters, and
what is the structure of the two gaps. From Eq. (\ref{gap_eq_toy})
one obtains the fourth-order equation for $L$-

\bea\label{L eqn} &&\frac{(u_{12} \alpha)^4}{4}L^4 -
\frac{(u_{12}\alpha)^2(u_{11} +
u_{22})}{2}L^3 + \\
&&\left( u_{11}u_{22}-u_{12}u_{12}'-(u_{12}\alpha)^2\right) L^2 +
(u_{11} + u_{22})L + 1=0\nonumber
\eea
where $u_{12}=N_{F_1}U_{12}$ and $u_{12}' = N_{F_2}U_{12}$. $T_c$
exists if this equation has a solution at $L >0$. The full
analysis of (\ref{L eqn}) is elementary but
cumbersome~\cite{aRG_wdFeSC} and it shows that for arbitrary
$U_{11}$, $U_{22}$, and $U_{12}$, the solution exists when the
parameters satisfy a certain inequality. However, for the case
relevant to \KFA, when $U_{11} = U_{22} = U_{12}$, i.e.,
$u_{11}u_{22} = u_{12}u_{12}'$ the solution exists for arbitrary
$N_{F_1}/N_{F_2}$, and $L$ is of order
 $1/(\sqrt{u_{12}u_{12}'}\alpha)$. We analyzed the gap structure
for this particular case and found that, predictably, the gaps on
the two pockets no longer satisfy $\Delta_{h_1} = - \Delta_{h_2}$
and also $r_1 \neq r_2$. As a result, in some range of
$N_{F_1}/N_{F_2}$, one gap remains nodal while in the other the
nodes are lifted. We illustrate this in Fig. \ref{fig:DOS}(b).

We see therefore that  nodal $s^{\pm}$ solutions are generally
favorable when $U_{11} \approx U_{22} \approx U_{12}$, which by
all accounts is the case for \KFA, and already infinitesimal
$\alpha$ gives rise to the nodes. The $T_c$ for a nodal solution
is lower that for a no-nodal solution (for which $T_c$ is
independent of $\alpha$ at small $\alpha$), what is also
consistent with the smaller $T_c$ in \KFA compared to optimally
doped Ba$_{1-x}$K$_x$Fe$_2$As$_2$.

\begin{figure}[t]
\includegraphics[width=3.8in]{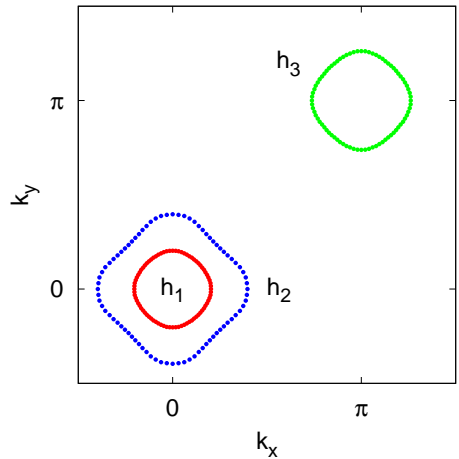}
\caption{\label{fig:fs} The FS for our 5-orbital model in the
unfolded BZ. Only hole FSs are present; the case corresponding to
\KFA. Two hole FSs, $h_1$ and $h_2$, are centered at the $\Gamma$
point, and one hole the FS, $h_3$, is centered at $(\pi,\pi)$. The
electronic structure was obtained in a band structure calculation
for the 5-orbital model\cite{LAHA_long}. The energy of the
$d_{3z^2-r^2}$ orbital was slightly modified to remove an
additional hole pocket at $(\pi,\pi)$.}
\end{figure}

\section{5-orbital model with 3 hole pockets}
\label{sec:full}

We now turn to a more microscopic description and solve the
pairing problem using as input the bare interactions $U_{h_i,h_j}
({\bf k}_{F_i}, {\bf k}_{F_j})$ obtained from the underlying
5-orbital model by converting Hubbard and Hund intra-orbital and
inter-orbital interactions into the band basis, i.e., dressing up
the interactions by angle-dependent coherence factors associated
with the transformation from orbital to band basis for electron
states~\cite{graser,LAHA_long}. We used band structure from
Ref.\onlinecite{LAHA_long} for the hole doping, when electron
pockets disappear leaving only two hole pockets at $\Gamma$ point
and one at $(\pi,\pi)$. The FS for this case is shown in Fig.
\ref{fig:fs}. We fit s-wave components of {\it all} interactions
by the first three angular harmonics: a constant, a $\cos 4
\theta_i$ term and a $\cos 8 \theta_i$ term, i.e, approximate
$U_{h_i,h_j} ({\bf k}_{F_i}, {\bf k}_{F_j})$  by

\bea\label{interactions} U_{h_i,h_j} ({\bf k}_{F_i}, {\bf
k}_{F_j}) &=& U_{ij}\left( 1+ \alpha_{ij}cos4\theta_k +
\alpha_{ji}cos4\theta_p +\right. \nonumber \\
&&\left. \beta_{ij}cos8\theta_k +
\beta_{ji}cos8\theta_p\right)
\eea

\begin{figure}[t]
$\begin{array}{cc}
\includegraphics[width=1.7in]{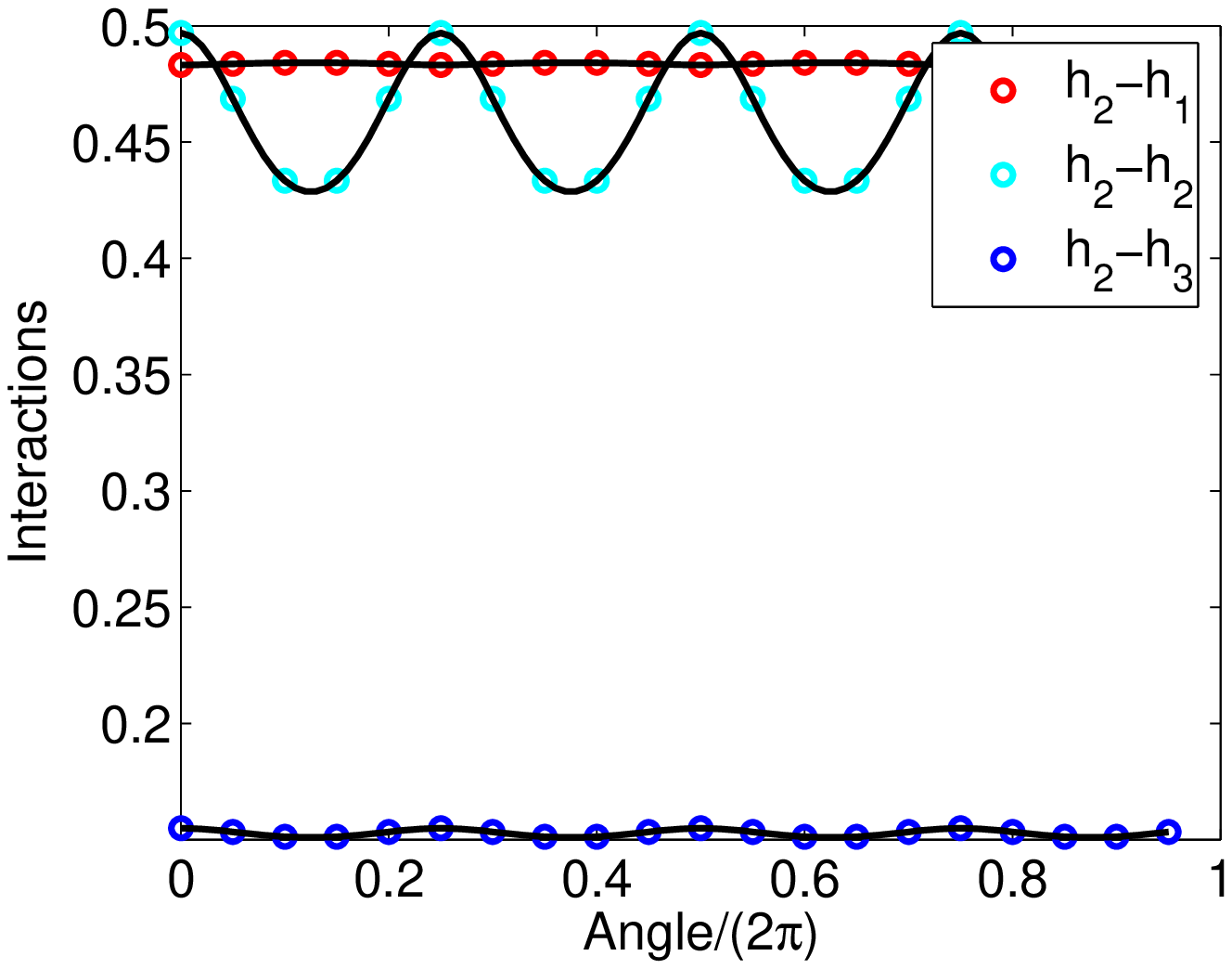}&
\includegraphics[width=1.7in]{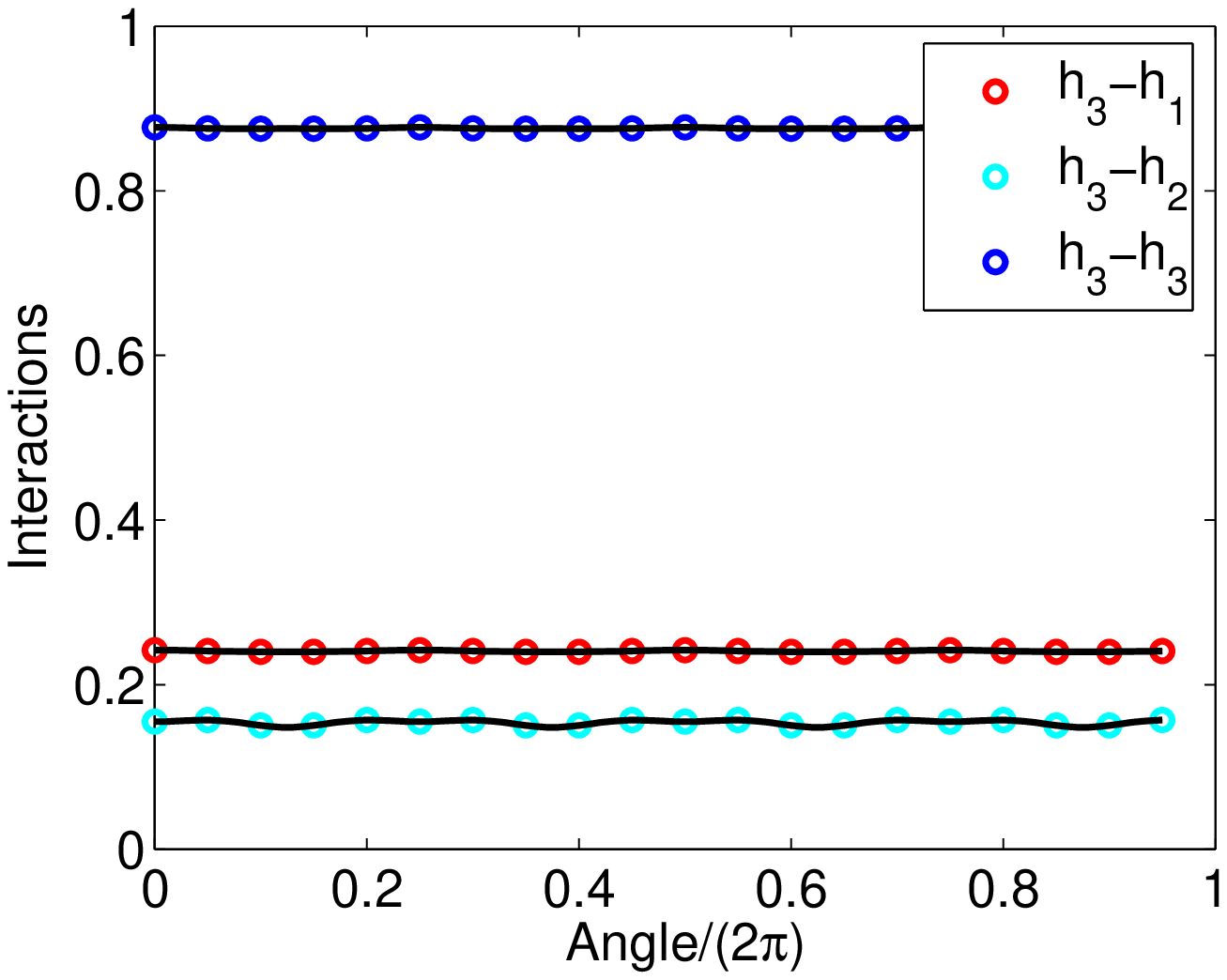}
\end{array}$
\caption{\label{fig:interaction fits} The fits of s-wave
components of the interactions obtained from 5-orbital model by
$U_{h_i,h_j} ({\bf k}_{F_i}, {\bf k}_{F_j})$. We set $\mu =-0.2$
(in the notations of Ref. \onlinecite{LAHA_long}), when electron
pockets just disappear.  We fix ${\bf k}_{F_i}$ to be along $x$
direction on the $h_2$ pocket(left) and $h_3$ pocket (right) and
vary ${\bf k}_{F_j}$ on all three hole FSs. The interactions are
almost perfectly reproduced by keeping angular harmonics up to
$8\theta$. }
\end{figure}

We show the fit in Fig. \ref{fig:interaction fits}. The agreement
is nearly perfect, which makes us believe that higher harmonics
can be safely neglected. The values of $U_{ij}$, $\alpha_{ij}$,
and $\beta_{ij}$ are shown in the Table \ref{tab}. Observe that
$U_{11}$, $U_{22}$ and $U_{12}$ are comparable, and the
angle-dependent parts of the interactions are rather small. These
are precisely the conditions we considered in the analytical
analysis above.

For the interactions given by Eq. (\ref{interactions}) each gap
$\Delta_{h_i}$  has the form

\bea\label{gaps} \Delta_{h_i}&=&\Delta_i \left(1 + r_i cos4\theta
+ {\bar r}_i cos8\theta \right) \eea

\begin{table*}[t]
\caption{$s$-wave interaction parameters obtained from fitting by
Eq. (\ref{interactions}) the interactions between hole pockets
obtained from 5-orbital model at strong hole doping (the case
corresponding to \KFA with $\mu =-0.20$, $U = 1$, $J = 0.25$, and
$V = 0.69$ in the notations of Ref.~\onlinecite{LAHA_long}). The
numbers in bold are the magnitude of the interactions involving
fermions from $h_{1}$ and $h_2$ pockets.} \label{tab}
\begin{ruledtabular}
\begin{tabular}{cccccccccccccc}
$U_{11}$&$\alpha_{11}$&$\beta_{11}$&$U_{22}$&$\alpha_{22}$&$\beta_{22}$&$U_{12}$
&$\alpha_{12}$&$\alpha_{21}$&$\beta_{12}$&$\beta_{21}$&$U_{33}$&$U_{13}$&$U_{23}$\\
\textbf{0.48}&0.00&0.00&\textbf{0.42}&0.08&0.00&\textbf{0.44}&0.00&0.08&0.00&0.00&
0.87&0.24&0.15\\
\end{tabular}
\end{ruledtabular}
\end{table*}

\begin{figure*}[t]
$\begin{array}{ccc}
\includegraphics[width=2in]{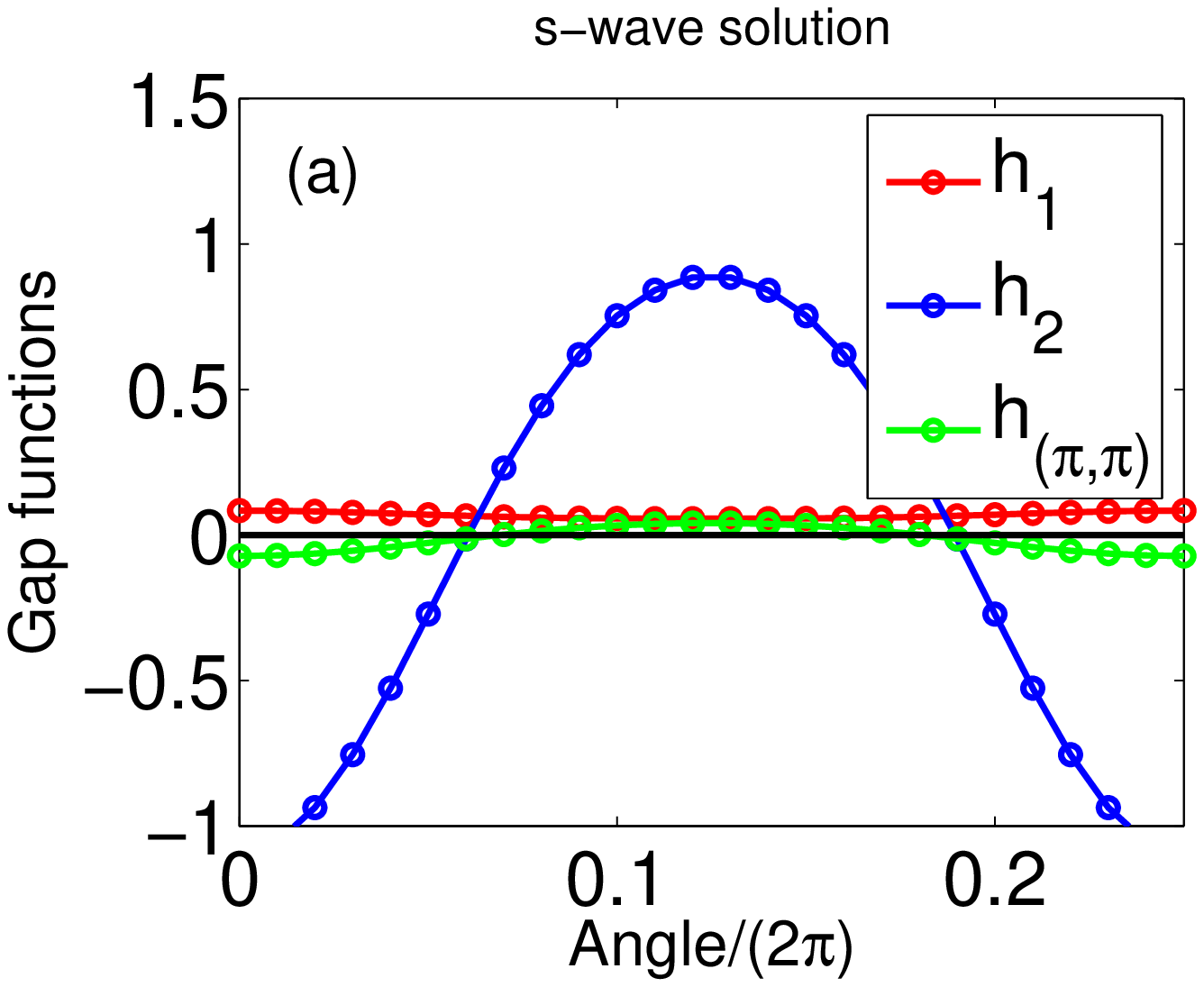}&
\includegraphics[width=2in]{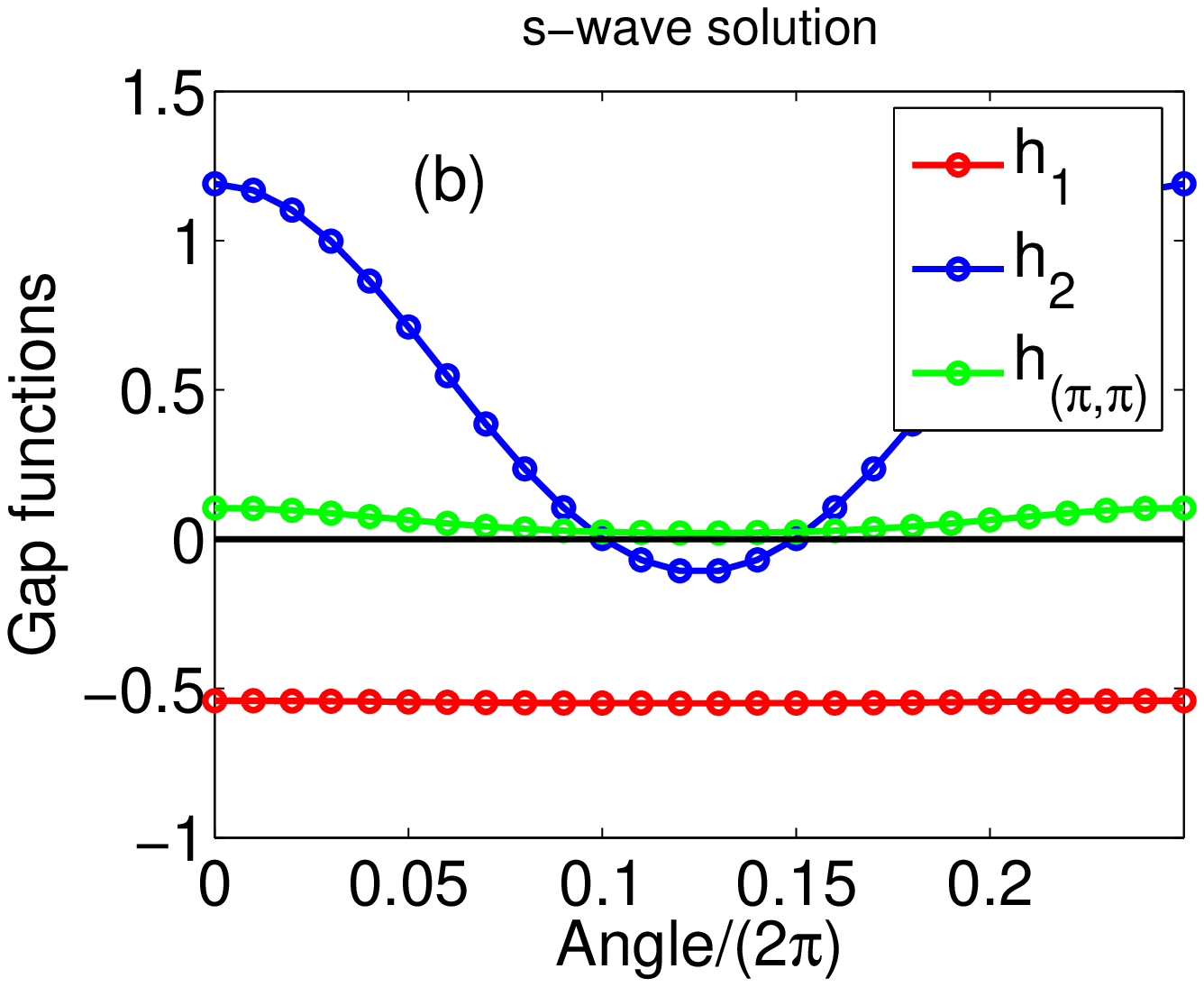}&
\includegraphics[width=2in]{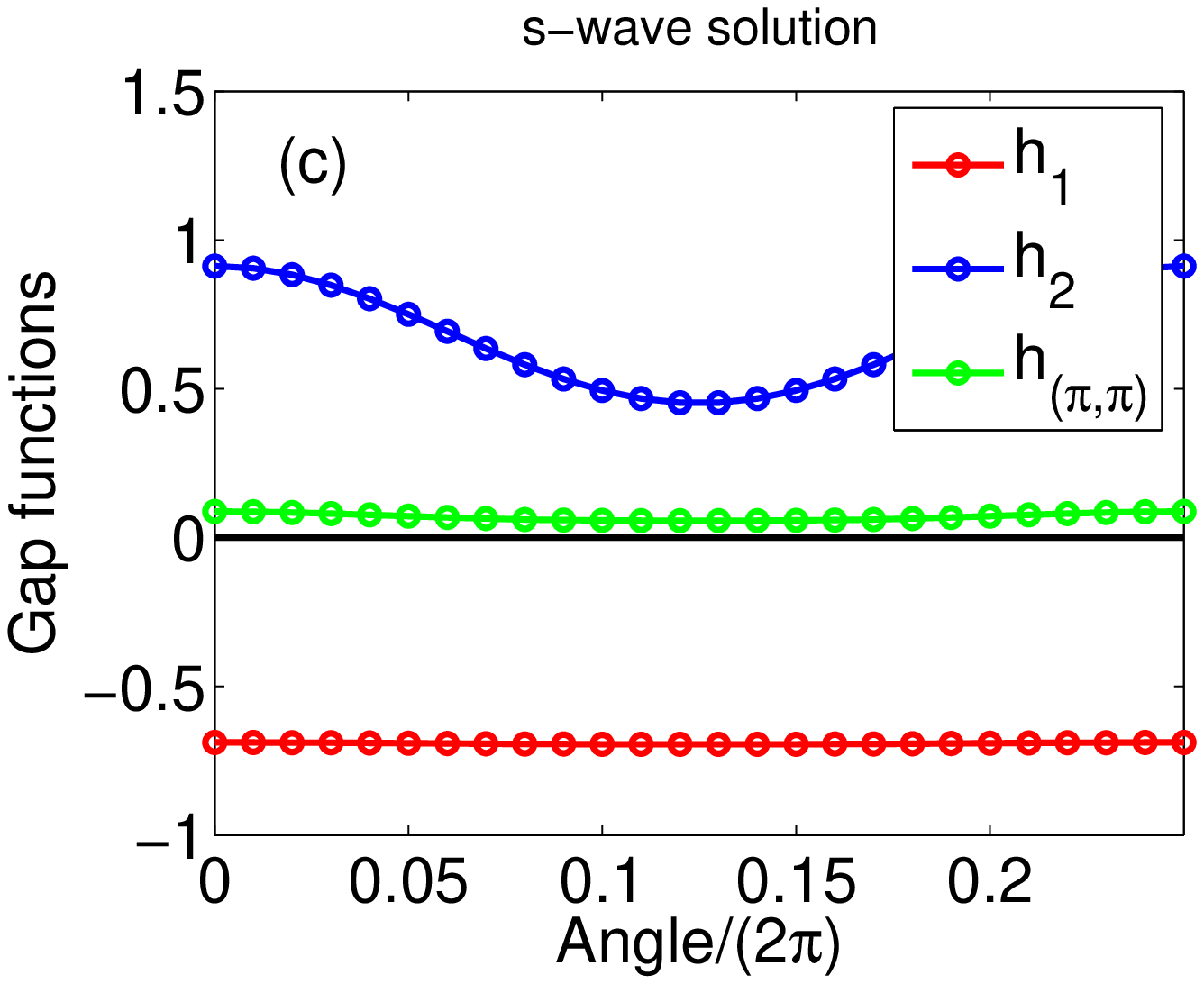}\\
\includegraphics[width=2in]{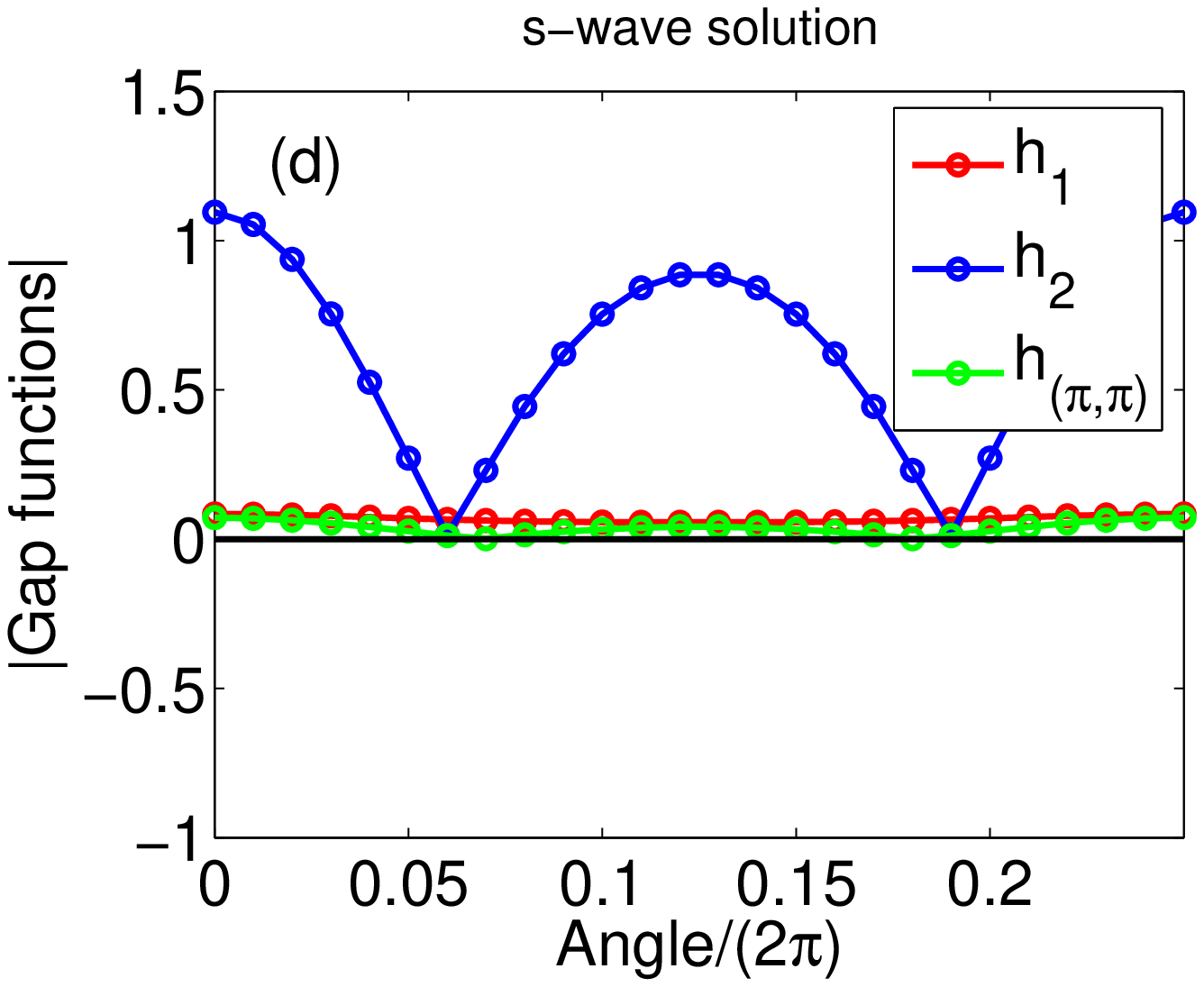}&
\includegraphics[width=2in]{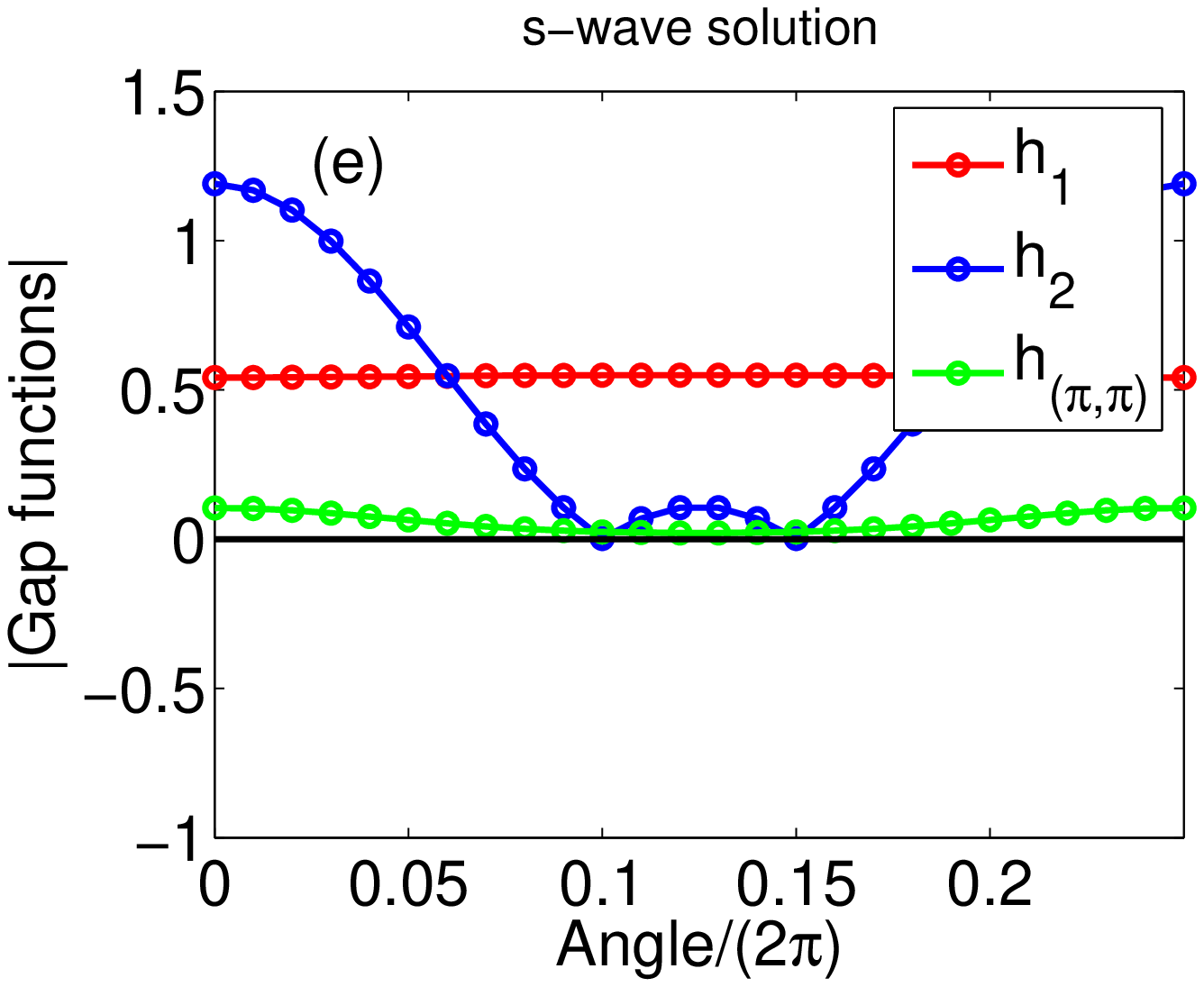}&
\includegraphics[width=2in]{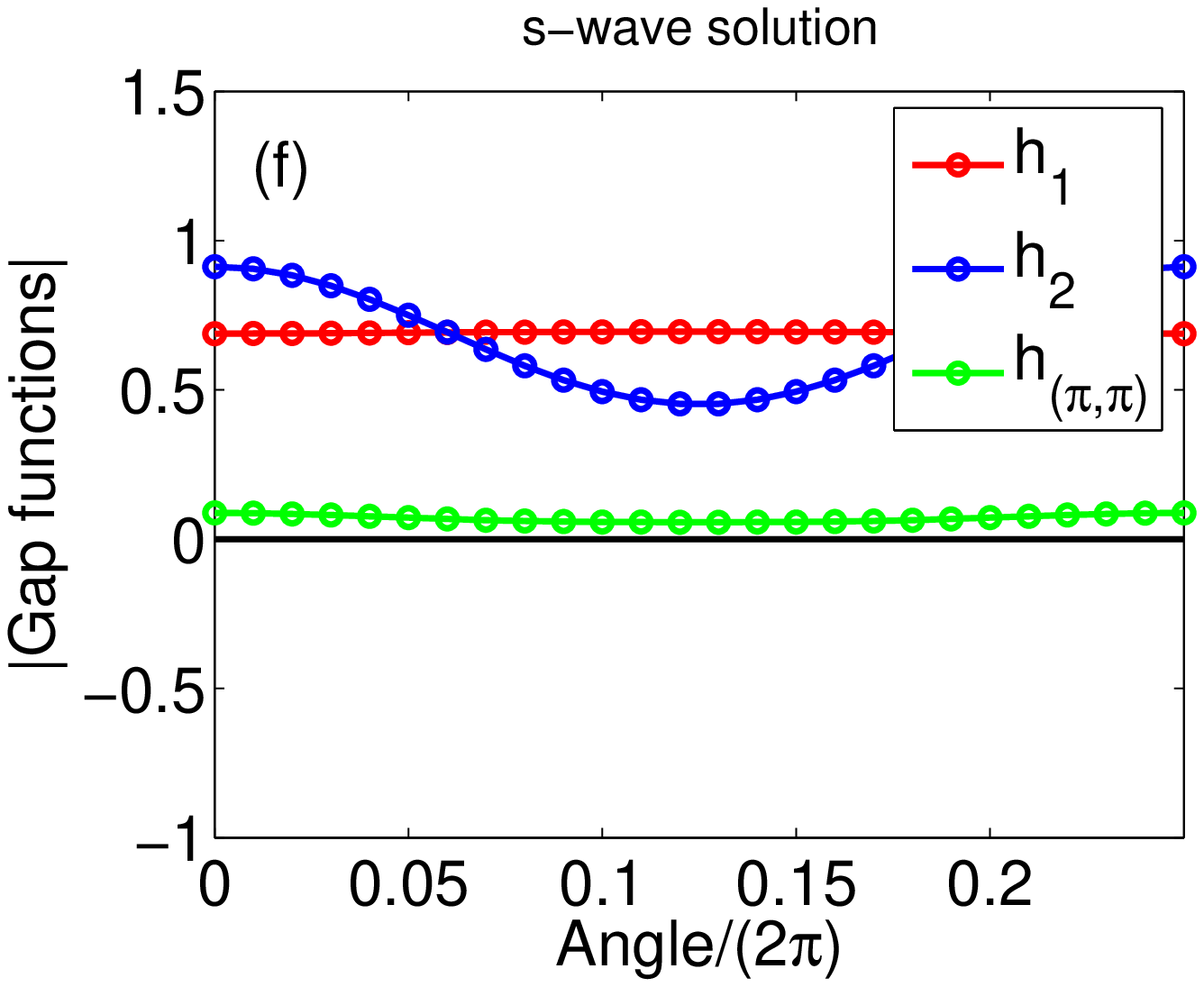}\\
\includegraphics[width=2.4in]{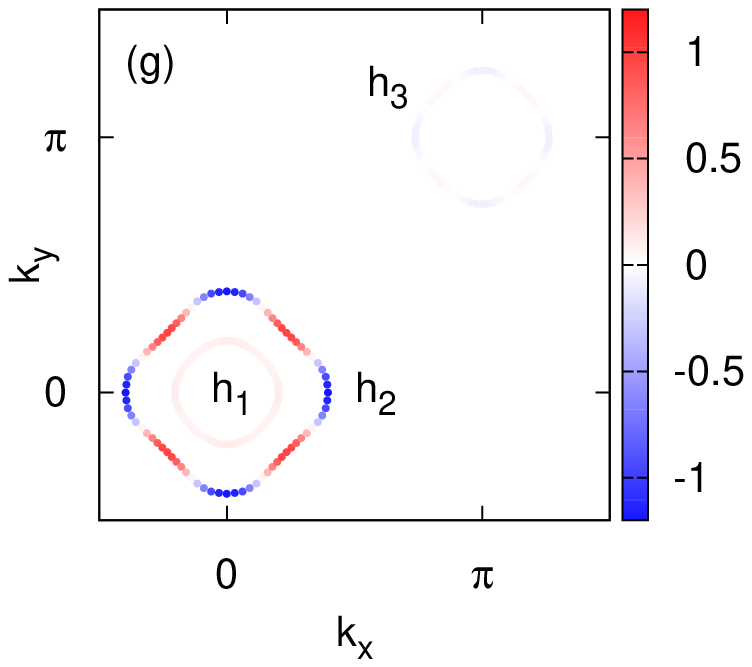}&
\includegraphics[width=2.4in]{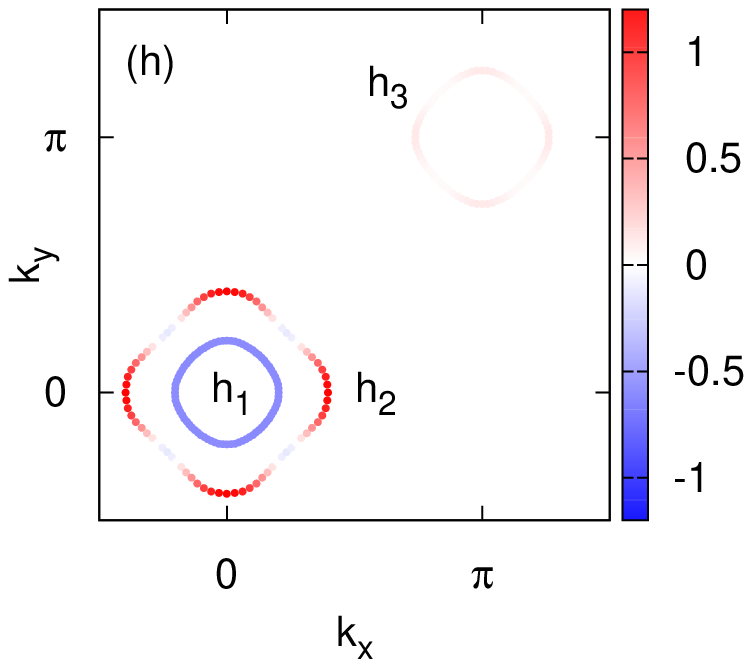}&
\includegraphics[width=2.4in]{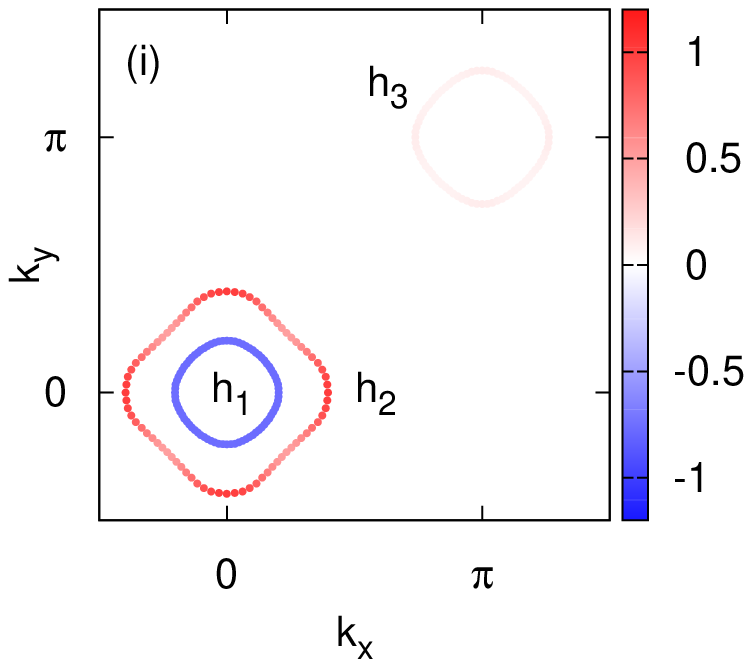}\\
\end{array}$
\caption{\label{fig:gaps3} (a) $s-$wave gap structure obtained by
solving $9\times9$ gap equation for a 5-orbital model. The
parameters are listed in Table \ref{tab}. (b) and (c) $s-$wave gap
structure obtained by changing $U_{11}\rightarrow 0.97U_{11}$ and
$0.95 U_{11}$, respectively. We see that the gap structure is very
sensitive to small changes in $U_{11}$. Panels  (d), (e), and (f)
--  the absolute values of the gaps from panels (a), (b), and (c),
respectively. Panels (g),(h) and (i) are the color-coded gap
structures shown on the entire FS which correspond to panels (a),
(b), (c) respectively.
The gap on $h_3$ FS is small and is barely visible on the color
plot.}
\end{figure*}

We  obtain $\Delta_i$ by solving $9\times 9$ matrix gap equation
and choosing the solution with the largest attractive eigenvalue.
For simplicity, we set all $N_{F_i}$ to be equal. The result is
presented in Fig. \ref{fig:gaps3}a.

The key observation in Fig. \ref{fig:gaps3}a is that $s-$wave gap
has nodes,  despite that angle-dependent terms in the interaction
potentials are quite small (see Table \ref{tab}).  We see that the
gap on the outer hole FS is the smallest, as we anticipated, i.e.,
the dominant interaction leading to $s^{\pm}$ superconductivity is
between $h_1$ and $h_2$ pockets. For the particular set of
interaction parameters used in 5-orbital model, the interplay
between the interactions is such that the gap on the $h_2$ FS is
the largest and has accidental nodes.  This is the consequence of
the fact that intra-pocket repulsion is larger on $h_1$ FS than on
$h_2$ FS. We verified, however, that the interplay between the gap
on $h_1$ and $h_2$ FSs depends on tiny details of the
interactions, and the two gaps become of comparable magnitude once
we change input parameters by a small amount. To illustrate this,
we show in  Fig. \ref{fig:gaps3}b,c the gap structure for the
cases when intra-pocket repulsion $U_{11}$ is changed by a tiny
bit -- to $0.97U_{11}$ and $0.95 U_{11}$, respectively. We see
that such a minor modification of the coupling strength
substantially affects the gap structure, making the magnitudes of
the gaps on $h_1$ and $h_2$ pockets comparable, and also lifting
the  nodes on the $h_2$ pocket.  This is fully consistent with our
argument that the angular dependencies of the gap on the $h_1$ and
$h_2$ pockets and the interplay between the magnitudes of these
two gaps  is the resonance phenomenon, which for small
$\alpha_{ij}$ crucially depends on the interplay between
$U^2_{12}$ and $U_{11} U_{22}$.

\section{The relative phase of  the $\cos 4 \theta$ components
of the gaps on different Fermi surfaces}
\label{sec:sign}

In this section we consider in some detail the issue about what
determines relative phases of $\cos 4\theta$ oscillations of the
s-wave gaps along different hole FSs not only in \KFA but also in
other FeSCs, which contain both hole and electron pockets. For the
latter, we also consider the interplay between $\cos 4\theta$ and
$\cos 2\theta$ gap variations along electron pockets. These issues
are relevant to experiments as angular variations of the gaps have
been detected not only in \KFA (Ref. \onlinecite{shin}) but also
in  LiFeAs (Refs. \onlinecite{sergei,ding_LiFeAs,rost}) in which
both hole and electron pockets are present. To make comparisons
with experiments more direct, we consider in this section the
folded BZ and measure $\theta$ for both hole and electron FSs as
deviations from the zone diagonal (the axis connecting hole and
electron pockets).

We found from our analytical analysis in Sec. \ref{sec:equiv} that
the sign of the $\cos 4\theta$ variation of the gap on $h_1$ and
$h_2$ FSs is the same as the sign of the corresponding component
of inter-pocket interaction [i.e., the sign of $r_{1,2}$ in Eq.
(\ref{gaps_toy})] is the same as the sign of $\alpha_{12,21}$ in
Eq. (\ref{interactions_toy}). We assumed that the signs of
$\alpha_{12}$ and $\alpha_{21}$ are positive and obtained
in-phase, $\cos 4 \theta$ oscillations on $h_1$ and $h_2$ FSs,
with gap maxima along the direction towards the would-be electron
pockets (see Fig. \ref{fig:DOS}). For the 5-orbital model,
$\alpha_{21} >0$ and $\alpha_{12}$ is essentially zero. One should
then expect $\cos 4\theta$ component of $\Delta_{h_1}$ to be small
and  $\cos 4\theta$ component of $\Delta_{h_2}$ to be positive,
i.e., $\Delta_{h_2}$ to be the largest in the direction towards
would be electron pockets.  This agrees with the actual solution
in Fig. \ref{fig:gaps3}. This form of $\Delta_{h_2}$ is consistent
with the laser-ARPES results for \KFA (Ref. \onlinecite{shin}),
although the data show two additional features not captured in our
analysis: (i) $\Delta_{h_1}$ also has a substantial angular
dependence, and (ii) $\cos 8 \theta$ gap modulations are
substantial for both $\Delta_{h_1}$ and $\Delta_{h_2}$.

The issue we address now is what determines the sign of
$\alpha_{ij}$.  We argue that two effects contribute, one is the
shape of the FS, another is the type of s-wave function which
gives the largest contribution to the pairing interaction.

\begin{figure}[t]
$\begin{array}{cc}
\includegraphics[width=1.7in]{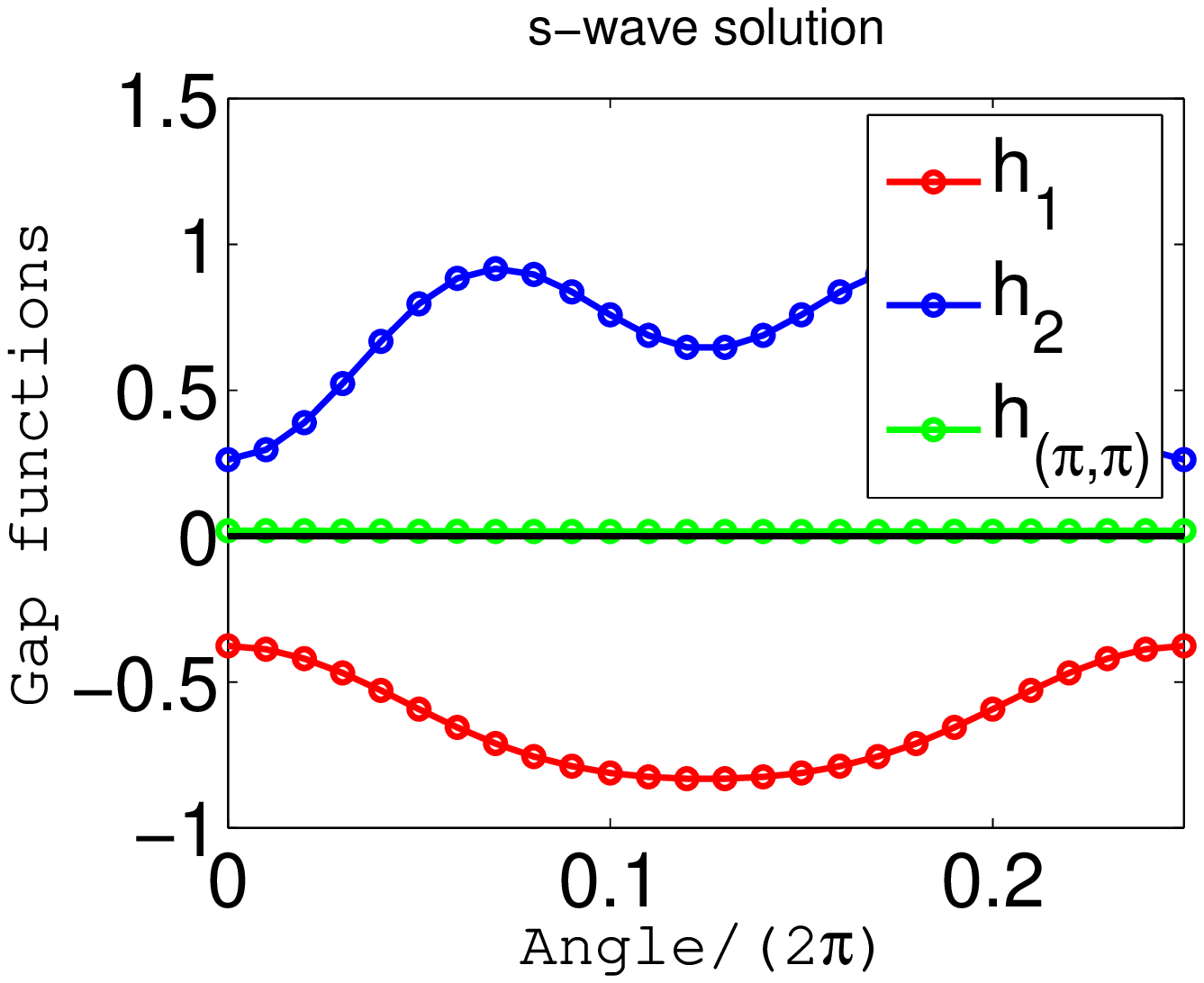}&
\includegraphics[width=1.7in]{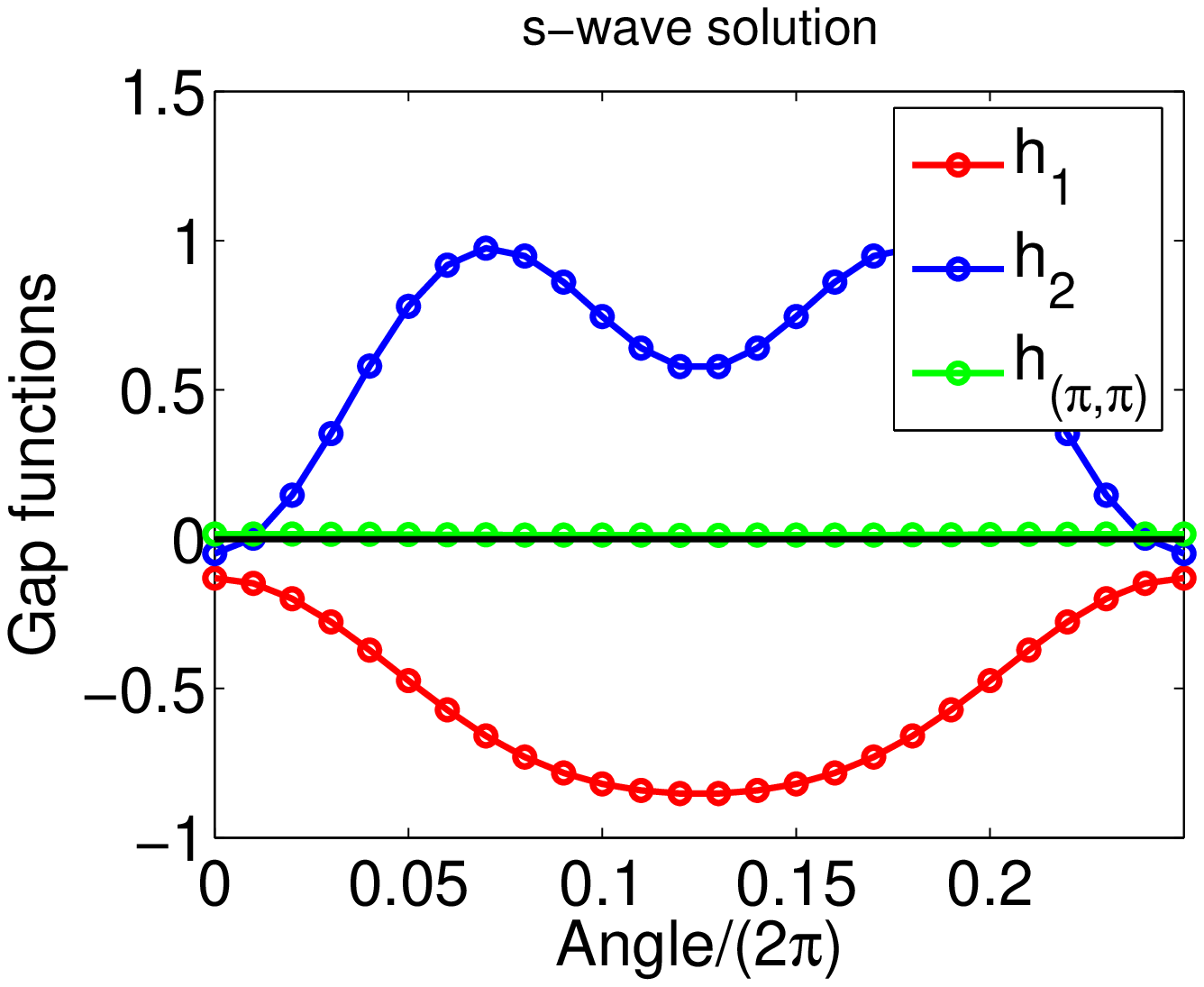}
\end{array}$
\caption{\label{fig:new} \emph{Left} -- $s-$wave gaps along the two
FSs $h_1$ and $h_2$ centered at $\Gamma$ point obtained by solving
$9\times 9$ gap equations with $U_{h_i,h_j} ({\bf k}_{F_i}, {\bf
k}_{F_j})$ obtained from the same 5-orbital model as before, but
with RPA-renormalization of the interactions. The primary effect
of RPA renormalization on hole-hole interaction is the increase of
the overall magnitude of $U_{ij}$. The increase is somewhat larger
for $U_{12}$ than for $U_{11}$ and $U_{22}$ simply because the
renormalized interaction is still larger for larger momentum
transfers.  \emph{Right} -- the same but with slightly larger
intra-pocket $U_{11,22} \rightarrow 1.23 U_{11,22}$ and slightly
smaller inter-pocket $U_{12} \rightarrow 0.85 U_{12}$, in which
case the gap on $h_{2}$ FS contains nodes. The main difference
between this figure and Fig. \ref{fig:gaps3} is $45^o$ rotation of
the direction of the gap maxima.}
\end{figure}

The reasoning goes as follows. Consider the pairing interaction
$U(k_1,k_2)$ between  two fermions located on the either $h_1$ or
$h_2$ FSs. The  $s-$wave component of the interaction is a
combination of products of the $A_{1g}$ eigenfunctions $\phi_i
(k_1) \phi_j (k_2)$, where (in the folded zone) $\phi_j (k) = 1$,
$\cos k_x + \cos k_y$, $\cos k_x \cos k_y$, etc. We consider
interactions between particles on the FS, hence ${\bf k_i} = {\bf
k}_{F,i}$. Because $k_1$ and $k_2$ are small, $\phi_i (k)$ can be
expanded in $k$. To the order $k^4$, the dependence is  $(1 - b_2
{\bf k}^2   + b_{4} {\bf k}^4  \cos 4 \theta) + ...$, where `...'
stand for subleading terms. For all eigenfunctions, $b_2>0$. And
$b_{4}$ is zero if $\phi_j (k) =1$, it is negative if $\phi_j (k)
= \cos k_x + \cos k_y$, $\cos 2k_x + \cos 2k_y$, etc, and it is
positive if $\phi_j (k) = \cos k_x \cos k_y, \cos 2k_x \cos 2k_y$,
etc. For a circular FS, ${\bf k}^2_F$ is just a constant and $\cos
4 \theta$ dependence comes exclusively from $b_4$ term. If a FS at
$\Gamma$ point is elongated (such that the deformation still
respects $C_4$ symmetry of the lattice), $k^2_F$ along this FS can
be modeled as ${\bf k}^2_F={\bf k}^2_{F_0} + \varepsilon_k~
cos4\theta$, where ${\bf k}^2_{F_0}$ is a constant and
$\varepsilon_k >0$  if the FS is elongated along $x$ and $y$
directions, and $\varepsilon_k < 0$ if the FS is elongated along
$x = \pm y$.  This gives rise to an additional $\cos 4 \theta$
dependence of the interaction potential, with the prefactor $- b_2
\varepsilon_k$. The combined prefactor of the $\cos 4 \theta_k$
term in $U_{ij} (k,p)$ is then $\alpha_{ij} \propto = -b_2
\varepsilon_k  +b_{4}$. The signs of $\alpha_{12}$ and
$\alpha_{21}$ then generally depend on the interplay between the
elongation of the FS and the type of the key $s-$wave
eigenfunction.  The same reasoning can be also applied to $h_3$
hole pocket.

The generic implication is that there is no simple way to relate
the sign of $\cos 4 \theta$ component of the gap to a FS geometry.
And the analysis based on 5-orbital model reflects the
non-universality: the maxima of the gaps on the FS $h_2$ are along
the direction of elongation of the FS (see Figs. \ref{fig:fs} and
\ref{fig:gaps3}) if we use bare interactions, as we did in the
previous section, but the maxima rotate by $45^o$ if we use
RPA-renormalized interactions. We show the results for the latter
case in Fig. \ref{fig:new}. In previous studies of 5-orbital model
for smaller hole dopings, when both hole and electron pockets are
present~\cite{RPA_wdFeSC_2,graser}, s-wave gaps on the FSs $h_1$
and $h_2$ have maxima along the direction of elongation of each of
the two FSs, but how sensitive the result is to the variation of
parameters is unclear.

The situation is more transparent in the case when $b_{4}$ is
small (e.g., when the leading eigenfunction is $\phi_j (k) = 1$).
Then the sign of a $\cos 4 \theta$ component of the gap $\Delta
(k)$ {\it anti-correlates} with the elongation of the
corresponding FS, i.e., if a FS is elongated along $x$ and $y$,
the gap on this FS has  maxima along $x = \pm y$ directions, and
if a FS is elongated along $x = \pm y$, the gap has maxima along
$x$ and $y$ directions.

This simple reasoning works amazingly well for the experimental
data.  Indeed, according to laser-ARPES data~\cite{shin}, the
measured FSs $h_1$ and $h_2$ are elongated along $x$ and $y$
directions, while  $\Delta_{h1}$ and $\Delta_{h2}$ extracted from
ARPES  have  $\cos 4 \theta$ terms with negative prefactors (the
behavior is the same as in Fig. \ref{fig:new}). It also works for
another Fe-pnictide superconductor, LiFeAs, in which the $\cos 4
\theta$ variation of the gaps on the hole FSs have been measured in ARPES
experiments~\cite{sergei,ding_LiFeAs} and extracted from the STM
data~\cite{rost}. The hole pockets around $\Gamma$ point  in
LiFeAs have been identified in ARPES
measurements~\cite{ARPES_LiFeAs_1} and extracted from de Haas-van
Alphen oscillations\cite{dHvA}. The two groups agree that the inner
hole pocket is either very small or doesn't even exist and that
outer hole pocket is elongated along $x = \pm y$. If the $\cos 4
\theta$ dependence is predominantly due to a FS elongation,  the
gap maxima on the $h_3$ pocket should then be at $45^o$ with
respect to the direction towards electron pockets.  This agrees
with both ARPES and STM results~\cite{sergei,rost}. There is a
discrepancy between ARPES and dHvA results concerning the size of
the middle hole pocket~\cite{comments} (it is larger in the
extraction from dHvA). Still, according to ARPES and STM
measurements~\cite{ARPES_LiFeAs_1,rost}, the middle pocket is
elongated along $x$ and $y$, hence the gap maxima on this FS
should be in the direction towards electron pockets (i.e., along
$x = \pm y$). The $\cos 4 \theta$ variation along $h_2$ FS has not
been directly measured in ARPES, but it was extracted from the
analysis of the STM data (Ref. \onlinecite{rost}) and found to be
along the direction towards electron pockets, again in agreement
with our simple reasoning. Whether the anti-correlation between
the direction of the elongation of a FS and the direction along
which s-wave gap is specific to \KFA and LiFeAs or is more
``universal" remains to be seen.

ARPES measurements on LiFeAs~\cite{sergei,ding_LiFeAs} also
detected variations of the gaps along the two electron FSs, which
in the folded zone are  inner and outer electron pockets centered
at $(\pi,\pi)$. These pockets are intersecting ellipses splitted
due to small hybridization. The hybridization vanishes (in the
absence of spin-orbit terms) at the crossing points, and, as the
consequence, inner and outer pockets touch each
other~\cite{sergei,ding_LiFeAs} along $k_x = \pi$ and $k_y=\pi$
lines (i.e., at $45^o$ with respect to the direction towards
$\Gamma$ point).

According to theory, $s-$wave gaps on electron FSs should have
both $\cos 4 \theta$ and $\cos 2 \theta$ variations. $\cos 4
\theta$ variations originate by the same reasons as on hole FSs,
i.e., from the expansion to forth order in $\pi-k$ of the $s$-wave gap
functions which do not vanish at $(\pi,\pi)$, like $\cos k_x +
\cos k_y$ or $\cos k_x \cos k_y$. By obvious reasons,  $\cos 4
\theta$ terms have the same signs on both electron FSs.  $\cos 2
\theta$ gap variations have different signs on the two electron
FSs (before hybridization) and originate by two reasons: (i)
electron pockets are ellipses, and (ii) s-wave gap function
generally contains components in the form $\cos (2n+1) k_x/2 \cos
(2n +1) k_y/2$  which vanish at $(\pi,\pi)$ and yield $\cos 2
\theta$ dependencies in the expansion near $(\pi,\pi$) (see e.g.,
Ref. \onlinecite{CVV}).  When hybridization is rather weak, the
$\pm \cos 2 \theta$ terms on the non-hybridized FSs become $\pm
|\cos 2 \theta|$ after hybridization, and the gaps on the inner
and outer FSs have angular dependence in the form

\bea\label{eq:2}
\Delta_{inner}(\theta)&=&\Delta_0 \left( 1+ r_2 |\cos2\theta| + r_4\cos4\theta \right) \nonumber\\
\Delta_{outer}(\theta)&=&\Delta_0 \left( 1- r_2 |\cos2\theta| +
r_4\cos4\theta  \right) \eea

For strongly elliptical FSs,
like in \BFCA,  $\cos 2 \theta$ terms should be dominant, and RPA
calculations for the band structure of \BFCA  do show that the
angular dependence of a gap along an electron FS is predominantly
$\cos 2 \theta$ ~\cite{RPA_wdFeSC_2,graser,LAHA_long,kuroki_2}.
However, electron FSs in LiFeAs are almost circular, in which case
it is likely that $\cos 2 \theta$ terms are comparable to $\cos 4
\theta$ terms.

\begin{figure}[t]
$\begin{array}{cc}
\includegraphics[width=1.7in]{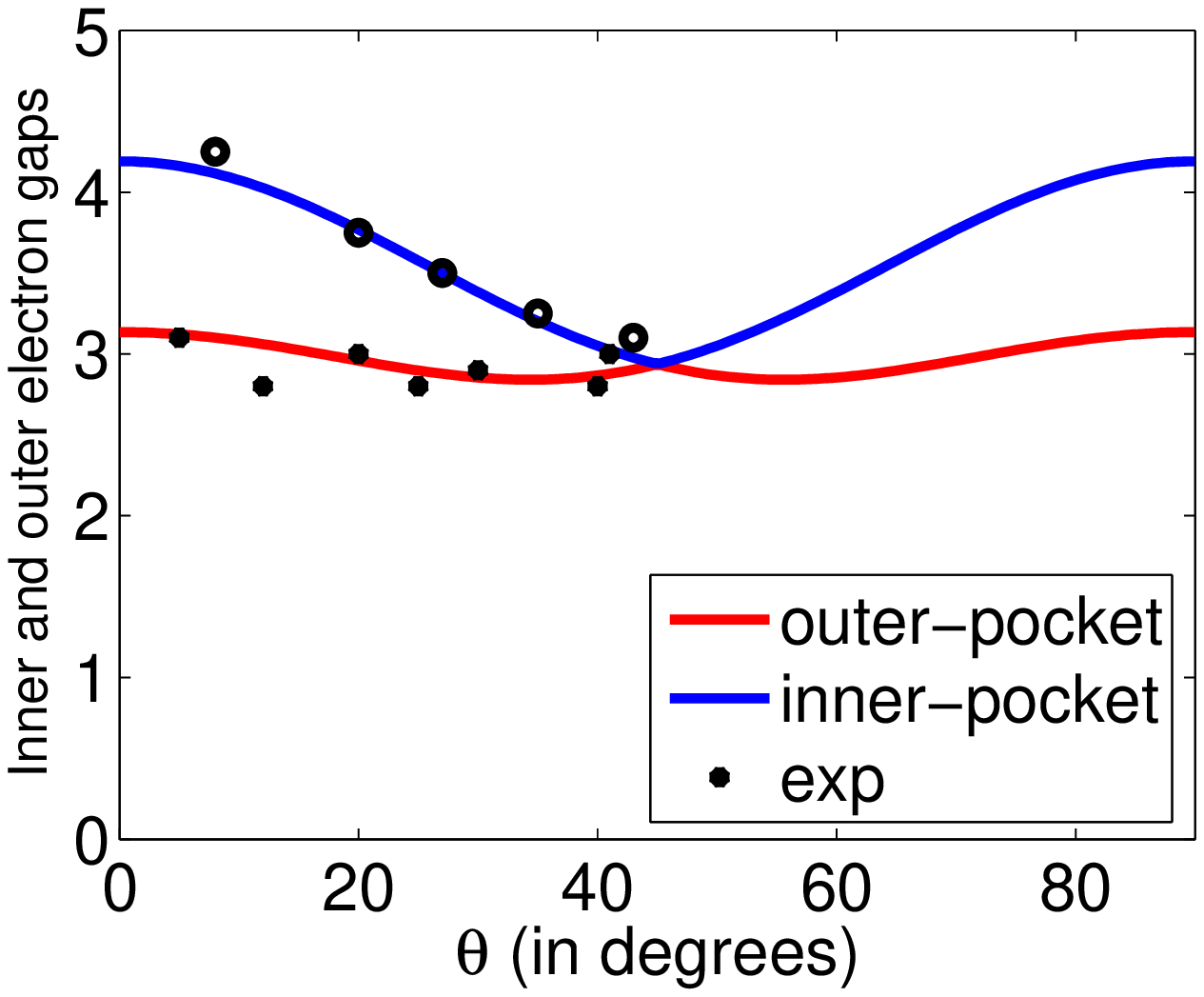}&
\includegraphics[width=1.7in]{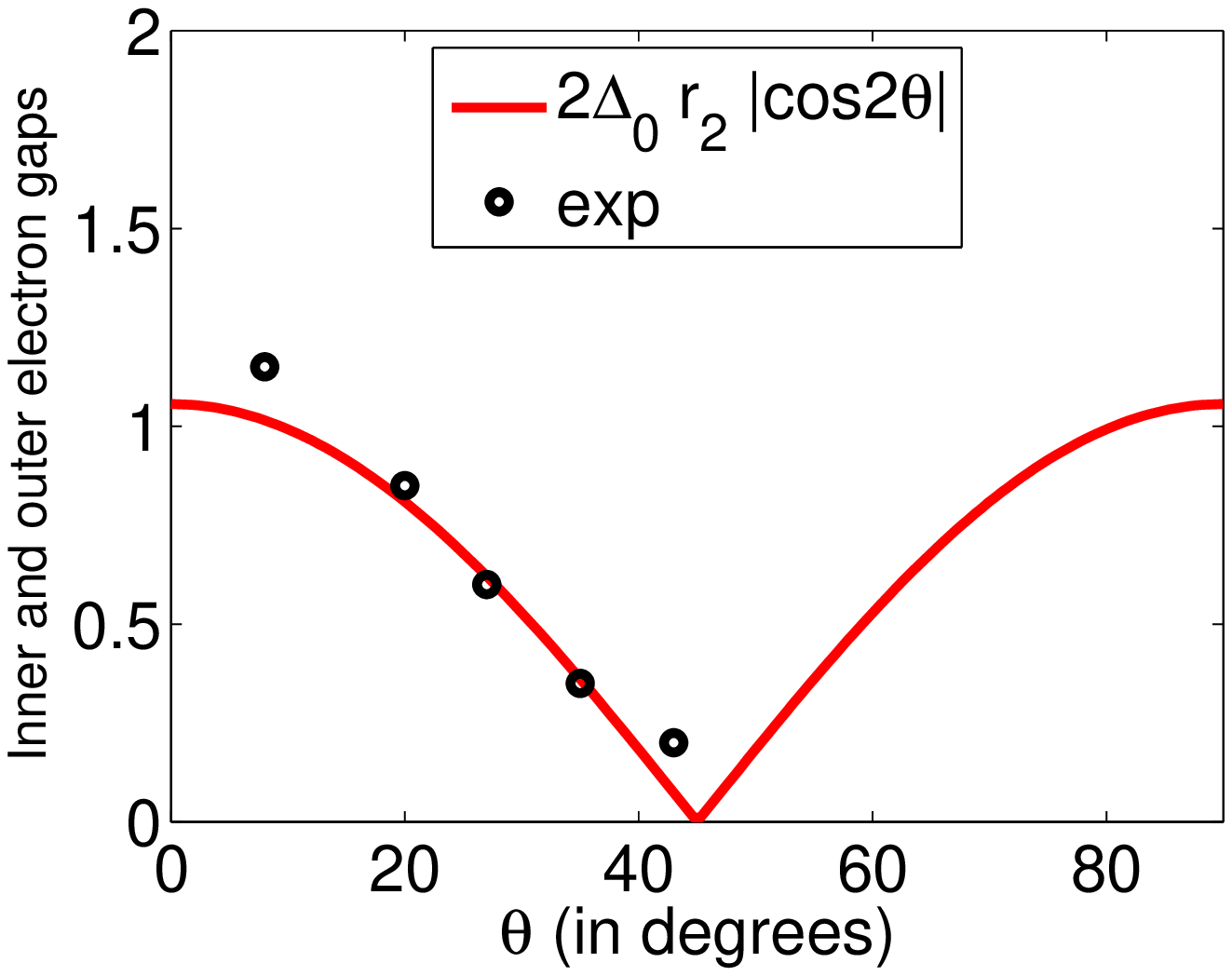}
\end{array}$
\caption{\label{fig:ding} \emph{Left} -- Fit, using
Eq.~\ref{eq:2}, to the gap structure on electron pockets measured
in Ref. \onlinecite{ding_LiFeAs}. The symbols are the experimental
data, the lines are theoretical curves. The parameters extracted
from the fit are $\Delta_0=3.3$meV, $r_2=0.16$, $r_4=0.11$.
\emph{Right} -- Plot of $\Delta_{inner}-\Delta_{outer}=2\Delta_0
r_2|\cos2\theta|$. The experimental data nicely fall onto
$\cos2\theta$ dependence.}
\end{figure}

ARPES experiments detect both $r_2$ and $r_4$ terms. In Fig.
\ref{fig:ding} we show a fit to the experimental data on the
electron gaps from Ref. \onlinecite{ding_LiFeAs} (the data from
Ref. \onlinecite{sergei} are very similar). We see that the
angular dependence of each of the two gaps is quite accurately
described by Eq. \ref{eq:2}, and the difference between the gaps
in the inner and outer pockets scales as $\cos 2\theta$, as it
should according to Eq. \ref{eq:2}. The parameters $r_2$ and $r_4$
extracted from the fit are $r_2 =0.16$ and $r_4 =0.11$.

Some useful information about underlying interaction can be also
obtained from the fact that both $r_2$ and $r_4$ are positive. For
near-circular electron pockets, the primary reason for $\cos 2
\theta$ variation is the dependence of the pairing interaction on
$\theta$ along the electron FS, which is generally in the form $
U(\theta) = U_0 \left(1+\beta cos2\theta+...\right)$. One can
easily make sure that the sign of $r_2$ is the same as the sign of
$\beta$. The latter, in turn depends on the orbital character of
FSs involved in the pairing interaction. If this interaction is
predominantly between hole and electron pockets, then it is
maximized in the direction towards the $\Gamma$ point
$(\theta=0)$~\cite{kordyuk} resulting in positive $\beta$. Then
$r_2$ should be positive as well, which is consistent with the
sign of $r_2$ extracted from the fit. This agreement suggests that
the electron-hole interaction, which is believed to determine
pairing properties of 1111 and 122 materials,  may also be the key
pairing interaction in LiFeAs.

The sign of $r_4$ is less universal feature. As we said, one can
obtain $\cos 4 \theta$ terms in the gaps on electron FSs by
expanding in $\pi-k_{x,y}$  s-wave eigenfunctions which do not
vanish at $(\pi,\pi)$. A simple exercise in trigonometry shows
that the sign of $r_4$ depends on the type of eigenfunction -- the
eigenfunction $\cos k_x + \cos k_y$ gives $r_4<0$, while the
eigenfunction $\cos k_x \cos k_y$ gives $r_4 >0$. The data show
that $r_4 >0$ [the gap maxima are along the direction towards
$\Gamma$ point], and from this perspective $\cos k_x \cos k_y$ is
more likely candidate. However, to address the issue why this
function and not $\cos k_x + \cos k_y$  is the primary gap
component one has to perform full microscopic analysis.

\section{Conclusions}

We argued in this paper that superconductivity in \KFA, which has
only hole pockets, can be $s-$wave with the nodes in the gap. We
have demonstrated that such a state appears quite naturally if the
dominant interaction between fermions is the one at small momentum
transfer. This is the case when the system is far from a spin or a
charge density-wave instability, and the interactions can be well
approximated by their bare values.  We argued that in this
situation the pairing chiefly comes from interactions between the
two hole pockets centered at $\Gamma$ point in the unfolded BZ.
When inter-pocket and intra-pocket interactions between these two
hole pockets are of near-equal strength, $s-$wave solution exists,
but the gap changes sign between the hole pockets and has nodes on
at least one of them due to the resonant enhancement of the
contribution to the s-wave gap from $\cos {4\theta}$ and $\cos
{8\theta}$ components of the interaction  The nodes are not
symmetry-related and are located at accidental $\theta$. This
$s^{\pm}$wave state with nodes is consistent with thermodynamic,
transport, and laser-based ARPES measurements of \KFA and is in
our view a viable candidate for the pairing state in this
material.

We also provided a simple explanation for the relative phases of
the $4\theta$ components of the gaps along hole FS by relating the
signs of the $\cos 4 \theta$ terms to the shapes of the hole FSs
and argued that this simple explanation is consistent with the
data for \KFA and also for LiFeAs. We argued that in LiFeAs ARPES
experiments also detected both  $\cos 2 \theta$ and  $\cos 4
\theta$ gap variations along the inner and outer electron FSs.

\section{Acknowledgements}
We are thankful to S. Borisenko, S. Davis, R. Fernandes, P.
Hirschfeld, I. Eremin, D. Evtushinsky, A. Kordyuk,  K. Kuroki, Y.
Matsuda, A. Rost, M. Vavilov, and V.  Zabolotnyy for useful
discussions. This work was supported by NSF-DMR-0906953 (SM and
AVC) and by Humboldt foundation (A.V.C.). M.M.K. is grateful for
support from RFBR (grant 09-02-00127), Presidium of RAS program
N5.7, FCP scientific and Research-and-Educational Personnel of
Innovative Russia for 2009-2013 (GK P891 and GK16.740.12.0731),
and President of Russia (grant MK-1683.2010.2).

\end{document}